\documentclass[aps,prd,twocolumn,superscriptaddress,nofootinbib]{revtex4-1}

\usepackage[utf8]{inputenc}
\usepackage{microtype}
\usepackage{multirow}
\usepackage{amsmath}
\usepackage{wasysym}
\usepackage[dvipsnames]{xcolor}
\usepackage{graphicx}
\usepackage{dcolumn}
\usepackage{bm}
\usepackage{xspace}
\usepackage{nicefrac}
\usepackage{units}
\usepackage{hyperref}
\hypersetup{
    pdfnewwindow=true,      
    colorlinks=true,       
    linkcolor=blue,          
    citecolor=blue,        
    filecolor=blue,      
    urlcolor=blue           
}

\newcommand{\uETp}{\ensuremath{\nicefrac{\mathrm{V}}{\mathrm{cm}}~\nicefrac{\textrm{K}}{\mathrm{mbar}}}\xspace}
\newcommand{\uEp}{\ensuremath{\nicefrac{\mathrm{V}}{\mathrm{cm}}\nicefrac{}{\mathrm{bar}}}\xspace}
\newcommand{\ukiloEp}{\ensuremath{\nicefrac{\mathrm{kV}}{\mathrm{cm}}\nicefrac{}{\mathrm{bar}}}\xspace}
\newcommand{\ukiloE}{\ensuremath{\nicefrac{\mathrm{kV}}{\mathrm{cm}}}\xspace}
\newcommand{\uDiff}{\ensuremath{\nicefrac{\mu\mathrm{m}}{\sqrt{\mathrm{cm}}}}\xspace}
\newcommand{\percm}{\ensuremath{\nicefrac{1}{\mathrm{cm}}}\xspace}
\newcommand{\uDrift}{\ensuremath{\nicefrac{\mathrm{cm}}{\mu\mathrm{s}}}\xspace}
\newcommand{\umob}{\ensuremath{\nicefrac{\mathrm{cm}^2}{\mathrm{V}}\nicefrac{}{\mu\mathrm{s}}}\xspace}

\newcommand{\hptpc}{\mbox{\textrm{HPTPC}}\xspace}

\newcommand{\magboltz}{\mbox{\texttt{\textsc{MagBoltz}}}\xspace}
\newcommand{\heed}{\mbox{\texttt{\textsc{Heed}}}\xspace}
\newcommand{\nie}{\ensuremath{\mathrm{N}_{\mathrm{I,e}}}\xspace}
\newcommand{\wi}{\ensuremath{\mathrm{W}_\mathrm{I}}\xspace}
\newcommand{\garfield}{\mbox{\texttt{\textsc{Garfield++}}}\xspace}

\newcommand{\micromegas}{\mbox{MicroMegas}\xspace}

\newcommand{\argon}{\text{Ar}\xspace}
\newcommand{\methane}{\ensuremath{\mathrm{CH}_4}\xspace}
\newcommand{\ethane}{\ensuremath{\mathrm{C}_2\mathrm{H}_{6}}\xspace}
\newcommand{\propane}{\ensuremath{\mathrm{C}_3\mathrm{H}_{8}}\xspace}
\newcommand{\isobutane}{\ensuremath{i\mathrm{C}_4\mathrm{H}_{10}}\xspace}
\newcommand{\butane}{\ensuremath{\mathrm{C}_4\mathrm{H}_{10}}\xspace}
\newcommand{\neopentane}{\ensuremath{\mathrm{C}(\mathrm{C}\mathrm{H}_{3})_{4}}\xspace}
\newcommand{\cphq}{\ensuremath{\mathrm{C}_x\mathrm{H}_y}\xspace}
\newcommand{\coo}{\ensuremath{\mathrm{CO}_2}\xspace}
\newcommand{\polys}{\ensuremath{(\textrm{C}_8\textrm{H}_8)_n}}
\newcommand{\tetrafluor}{\ensuremath{\mathrm{CF}_4}\xspace}

\newcommand{\pten}{90\%~\textrm{Ar} + 10\%~\methane} 
\newcommand{\ppten}{$\textrm{P-10}$\xspace}
\newcommand{\pft}{50\%~\textrm{Ar} + 50\%~\methane}
\newcommand{\ppft}{$\textrm{P-50}$\xspace}
\newcommand{\pnty}{10\%~\textrm{Ar} + 90\%~\methane}
\newcommand{\ppnty}{$\textrm{P-90}$\xspace}
\newcommand{\alicegas}{90\%~\textrm{Ne} + 10\%~\coo}
\newcommand{\ttokgas}{95\%~\textrm{Ar} + 3\%~\tetrafluor + 2\%~\isobutane}
\newcommand{\propft}{50\%~\textrm{Ar} + 50\%~\propane}
\newcommand{\propnt}{5\%~\textrm{Ar} + 95\%~\propane}

\newcommand{\maxpropane}{95}
\newcommand{\maxisobutane}{35}

\newcommand{\pres}{\ensuremath{P}\xspace}
\newcommand{\temp}{\ensuremath{T}\xspace}

\newcommand{\dEdx}{\ensuremath{\nicefrac{\textrm{d}\textit{E}}{\textrm{d}\textit{x}}}\xspace}
\newcommand{\ETp}{\ensuremath{\nicefrac{E\temp}{\pres}}\xspace}
\newcommand{\Ep}{\ensuremath{\nicefrac{E}{\pres}}\xspace}

\newcommand{\vd}{\ensuremath{v_\mathrm{d}}\xspace}
\newcommand{\ldrift}{\ensuremath{l_\mathrm{d}}\xspace}
\newcommand{\dl}{\ensuremath{\sigma_\mathrm{L}}\xspace}
\newcommand{\dt}{\ensuremath{\sigma_\mathrm{T}}\xspace}
\newcommand{\dlt}{\ensuremath{\sigma_\mathrm{L,T}}\xspace}

\newcommand{\sbr}{S/B}
\newcommand{\proton}{\textrm{p}}

\newcommand{\oom}{\mathcal{O}}
\newcommand{\mob}{\mu}
\newcommand{\gevc}{\ensuremath{\textrm{GeV}/\textit{c}}\xspace}
\newcommand{\mevc}{\ensuremath{\textrm{MeV}/\textit{c}}\xspace}
\newcommand{\dptt}{\ensuremath{\delta\textit{p}_\textrm{TT}}\xspace}
\newcommand{\sdptt}{\ensuremath{\dptt^\textrm{smeared}}\xspace}
\newcommand{\pt}{\ensuremath{\textit{p}_\textrm{T}}\xspace}
\newcommand{\penningprob}{\ensuremath{\mathcal{R}}\xspace}

\newcommand{\comment}[1]{} 

\begin{document}

\title{Neutrino-hydrogen interactions with a high-pressure time projection chamber}

\author{Philip Hamacher-Baumann}
\email{hamacher.baumann@physik.rwth-aachen.de}
\affiliation{III.\ Physikalisches Institut, RWTH Aachen University, 52056 Aachen, Germany}

\author{Xianguo Lu}
\email{xianguo.lu@physics.ox.ac.uk}
\affiliation{Department of Physics, University of Oxford, Oxford OX1 3PU, United Kingdom}

\author{Justo Mart\'in-Albo}
\email{justo.martin-albo@ific.uv.es}
\affiliation{Instituto de F\'isica Corpuscular (IFIC), Universitat de Val\`encia \& CSIC, 46980 Valencia, Spain}

\date{\today}

\begin{abstract}
We investigate the idea of detecting pure neutrino-hydrogen interactions in a multinuclear target using the transverse kinematic imbalance technique [Lu \textit{et al.}, \href{https://journals.aps.org/prd/abstract/10.1103/PhysRevD.92.051302}{Phys.\ Rev.\ D\textbf{92}, 051302 (2015)}] in a high-pressure time projection chamber (\hptpc). With full solid-angle acceptance, MeV-level proton tracking threshold, state-of-the-art tracking resolution, and an $\oom(\unit[100]{m^3})$ gas volume at \unit[10]{bar}, an \hptpc could provide an opportunity to realize this technique. We propose the use of hydrogen-rich gases in the TPC to achieve high detection purity with large hydrogen mass. With the projected neutrino beam exposure at the DUNE experiment, neutrino-hydrogen events of the order of $10^4$ per year with purity above \unit[90]{\%} could be achieved with such an \hptpc using methane gas. In this paper, we present a systematic study of the event rate and purity for a variety of argon-alkane mixtures, and examine these gas candidates for the TPC tracking-related properties. 

\end{abstract}

\maketitle

\section{Introduction} \label{sec:intro}

\begin{figure}[tb]
\centering
\includegraphics[width=\columnwidth]{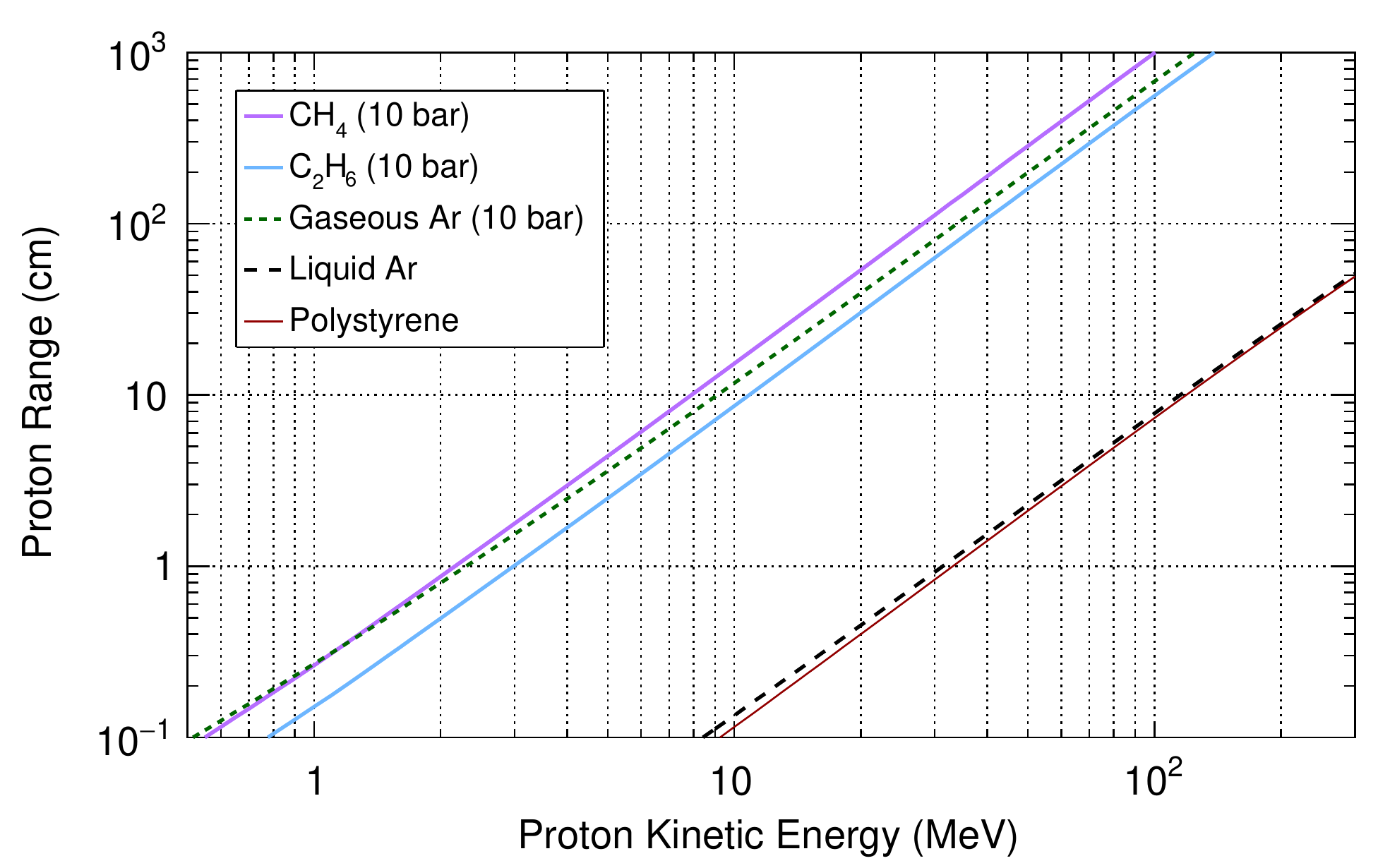} 
\caption{Calculated range (using the SRIM program~\cite{2010NIMPB.268.1818Z}) of protons in gaseous methane, ethane and argon at \unit[10]{bar} and \unit[25]{$^\circ$C}, as well as in liquid argon and polystyrene, as a function of the kinetic energy. Note the power-law dependence:   $\textrm{range}\sim(\textrm{kinetic energy})^2/\textrm{(material density)}$.  
}
\label{fig:range5}
\end{figure}

Neutrino-oscillation measurements rely on our understanding of neutrino interactions in the GeV regime to infer the neutrino energy and flux. To achieve the required interaction rates, neutrino detectors use materials that can be practically scaled up, like water, plastics or liquid argon, at the cost of dealing with complex neutrino-nucleus interactions that result in a major source of systematic uncertainties \cite{Alvarez-Ruso:2017oui, Betancourt:2018bpu}. Understanding neutrino interactions has become crucial for T2K~\cite{Abe:2019vii} and NOvA~\cite{Acero:2019ksn}, and so will be for DUNE~\cite{Abi:2020wmh} and Hyper-Kamiokande~\cite{Abe:2018uyc}. As intranuclear effects can only be inferred from final-state particles, detectors in future experiments are being designed for lower detection thresholds and larger acceptance.  One such detector concept is the high-pressure time projection chamber (\hptpc), which is one of the components of the future DUNE near detector~\cite{Abi:2020wmh}. The DUNE \hptpc, as currently envisioned, consists of a cylindrical detection volume of about \unit[100]{m$^3$} (about \unit[5.2]{m} in diameter and \unit[5]{m} in length) that holds a pressurized gas at \unit[10]{bar} and room temperature. To provide constraints on neutrino interactions at the DUNE far detector, which uses liquid argon, the default gas mixture of the \hptpc is \ppten (\pten), providing argon mass of about \unit[1.5]{t}. Housed in a magnet with a field strength of \unit[0.5]{T}, this \hptpc provides tracking and charge separation for particles   originating from the neutrino-gas interactions. In addition to the full (4$\pi$ solid angle) acceptance, its proton tracking threshold is \unit[3]{MeV} of kinetic energy~\cite{Abi:2020wmh}, over an order of magnitude smaller than in solid or liquid detectors (see Figure~\ref{fig:range5}). 

Due to the absence of nuclear effects in neutrino-hydrogen interactions, hydrogen would be the ideal target for a neutrino oscillation experiment, if it were possible to use it in large quantities without the presence of other nuclides. A hydrogen target would significantly improve the reconstruction of the neutrino energy spectra that is crucial for measuring neutrino oscillation probabilities~\cite{Lu:2015hea}. While hydrogen bubble chambers were used to detect neutrinos before the 1990s, in more recent experiments using plastic scintillator, mineral oil, or water as target, neutrino-hydrogen interactions are inseparable from carbon and oxygen background interactions (for a review, see, for example, Ref.~\cite{Tanabashi:2018oca}). Recently, it has been proposed~\cite{Lu:2015hea} that neutrino-hydrogen interactions from a neutrino beam could be selected in an event-by-event basis from a compound target that contains hydrogen if sufficient momentum resolution is achieved. The idea was to use the transverse kinematic imbalance (TKI) of the final-state particles with respect to the neutrino beam direction: with perfect tracking, interactions on hydrogen would have balanced final-state transverse momenta (that is, zero TKI) while the TKI on heavy nuclei is irreducibly wide due to  nuclear effects such as Fermi motion and final-state interactions (FSIs). As the original method has been discussed in Ref.~\cite{Lu:2015hea}, the focus of this work is the impact of the hydrogen amount in the target and the tracking performance that all together help suppress the nuclear background.

A large \hptpc, with hydrogen in its gas mixture, could be the ideal detector to realize this technique and provide high quality data on neutrino-hydrogen interactions. The default \ppten gas of the DUNE \hptpc contains only very limited hydrogen mass, and the  background from both carbon and argon is overwhelming. However, a TPC has the unique advantage of being flexible in switching the gas, the target material that neutrinos interact on. In this work, we discuss the feasibility of hydrogen-rich gas mixtures in a \hptpc, with a focus on the perspective of measuring neutrino-hydrogen interactions given the state-of-the-art tracking performance,  an $\oom(\unit[100]{m^3})$ gas volume at \unit[10]{bar}, and a neutrino rate as expected at DUNE. 

This paper is organized as follows: In Section~\ref{sec:tki} we review the TKI technique that allows the use of hydrogen-containing chemical compounds for pure neutrino-hydrogen interactions. In Section~\ref{sec:method} we introduce the \hptpc gas mixture candidates, examining them in terms of their hydrogen mass and purity. Because the TKI technique relies on the TPC tracking, we discuss the gas mixture properties in terms of drift velocity, diffusion, and gas gain in Section~\ref{sec:driftgas}. In Section~\ref{sec:summary} we summarize this study and discuss the outlook towards a full realization of measuring neutrino-hydrogen interactions with a \hptpc.

\section{The Method of TKI}\label{sec:tki}
In neutrino interactions on  nuclei other than hydrogen, the nuclear remnant carries away energy and momentum. The kinematics between  the incoming  neutrino and the outgoing particles are therefore imbalanced. If the neutrino energy is unknown, only the imbalance among the momenta transverse to the neutrino direction is experimentally accessible. The method of TKI uses the details of this imbalance to precisely identify intranuclear dynamics~\cite{Lu:2015tcr, Furmanski:2016wqo, Abe:2018pwo, Dolan:2018sbb, Lu:2018stk, Dolan:2018zye, Lu:2019nmf, Harewood:2019rzy, Cai:2019jzk, Cai:2019hpx, Coplowe:2020yea} or the absence thereof~\cite{Lu:2015hea, Duyang:2018lpe, Duyang:2019prb, Munteanu:2019llq}. In order to observe the balanced transverse momenta on hydrogen, all final-state particles need to be measured. While a gaseous TPC has the optimal  acceptance and detection  threshold for interactions on its gas, it is only sensitive to charged particles. Therefore, the particular neutrino-hydrogen interaction channels to consider are the ones with only charged final states, which dominantly are the following three-track events~\cite{Lu:2015hea} (for similar ideas, cf.\ Refs.~\cite{Duyang:2018lpe, Duyang:2019prb}; for antineutrino-hydrogen quasielastic interactions with a neutron in the final state, see Ref.~\cite{Munteanu:2019llq}):
\begin{align}
\nu+\proton&\rightarrow\mu^-+\proton+\pi^+,~\textrm{and}\label{eq:nures}\\
\bar{\nu}+\proton&\rightarrow\mu^++\proton+\pi^-,\label{eq:nubarres}
\end{align}
where $\nu$ and $\bar{\nu}$ are, respectively, a neutrino and an antineutrino, and $\mu$, $\proton$,  and $\pi$ are a muon, proton and pion, respectively. The $\proton\pi^+$ channel takes place primarily   through the Delta resonance $\Delta^{++}(1232)$ production, while for   $\proton\pi^-$, in addition to the Delta resonance $\Delta^0(1232)$, charge-neutral nucleon resonances with higher mass also contribute significantly (see discussions below).

\begin{figure}[!tb]
    \centering
    \includegraphics[width=\columnwidth]{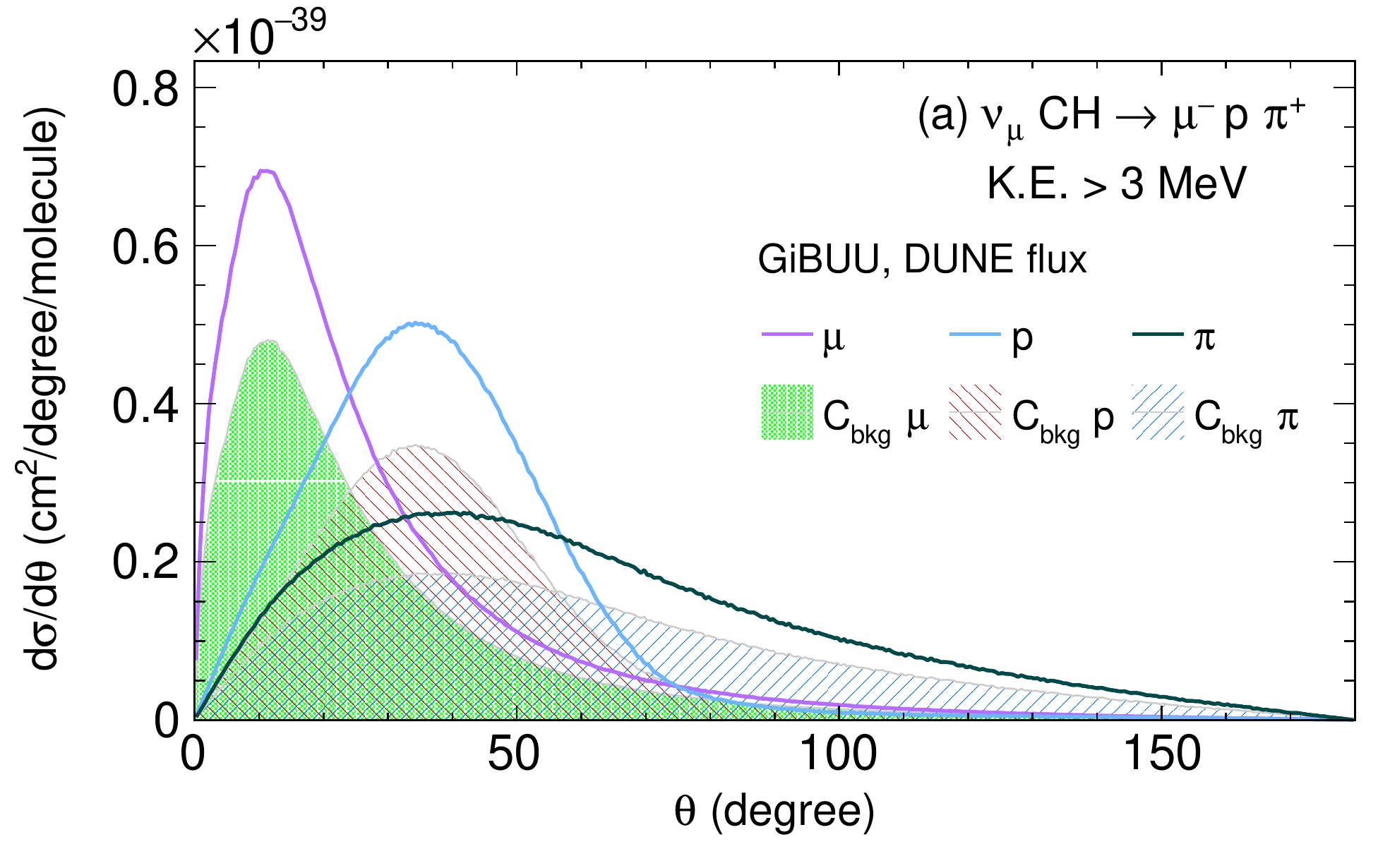}  
    \includegraphics[width=\columnwidth]{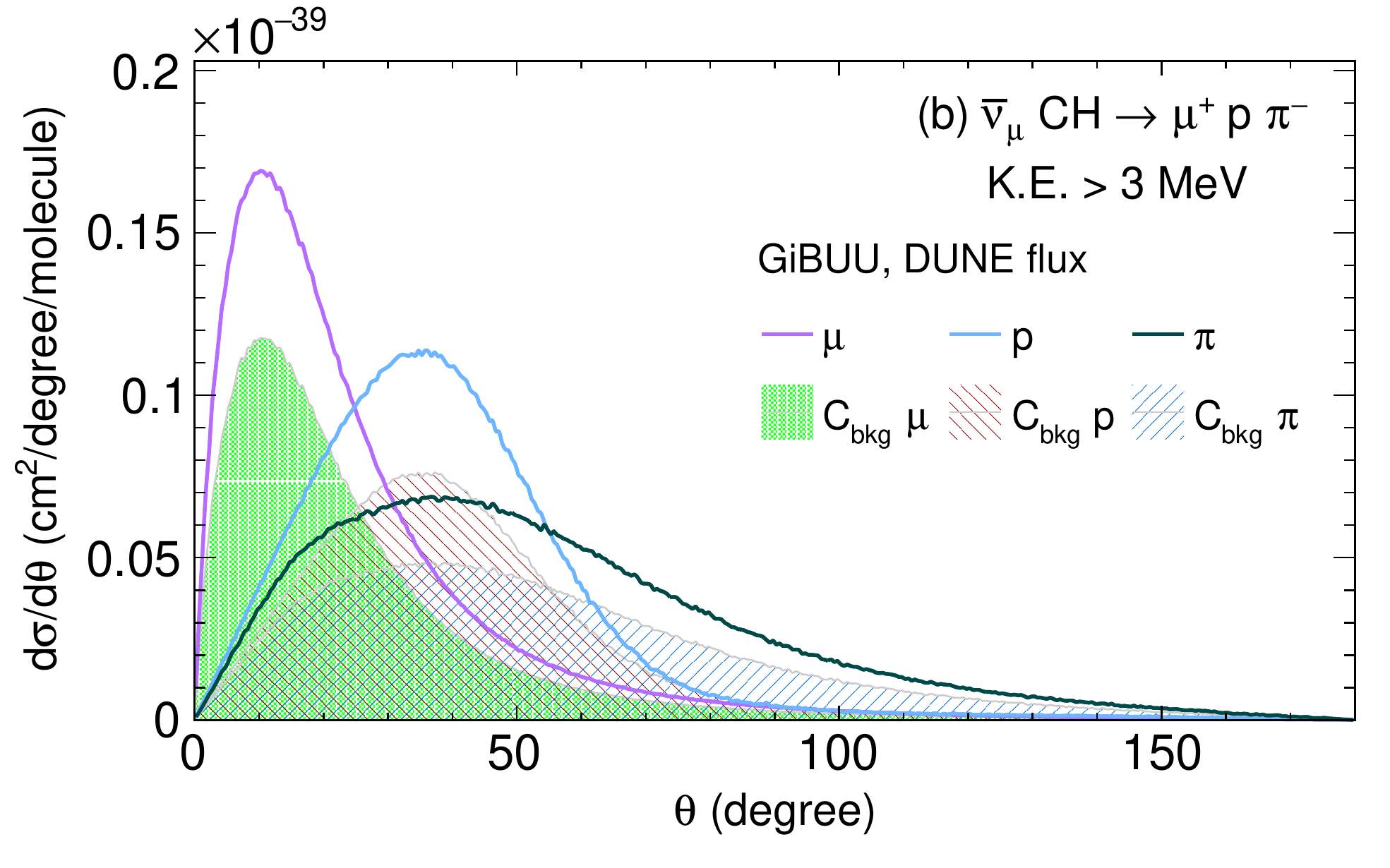}  
    \caption{Flux-averaged differential cross section as a function of the final-state particle polar angle $\theta$ with respect to the incoming neutrino (top) or antineutrino (bottom) interacting on a CH ``molecule". The cross section is calculated using GiBUU~\cite{Buss:2011mx}  with the DUNE fluxes~\cite{Abi:2020wmh}. The respective signal channels are Eqs.~(\ref{eq:nures}) and~(\ref{eq:nubarres}) where the kinetic energy (K.E.) of each final-state particle is greater than   \unit[3]{MeV}.  Comparison is made among  the final-state particles. The carbon backgrounds are shown as shaded histograms.}
    \label{fig:theta}
\end{figure}

\begin{figure}[t]
    \centering
    \includegraphics[width=\columnwidth]{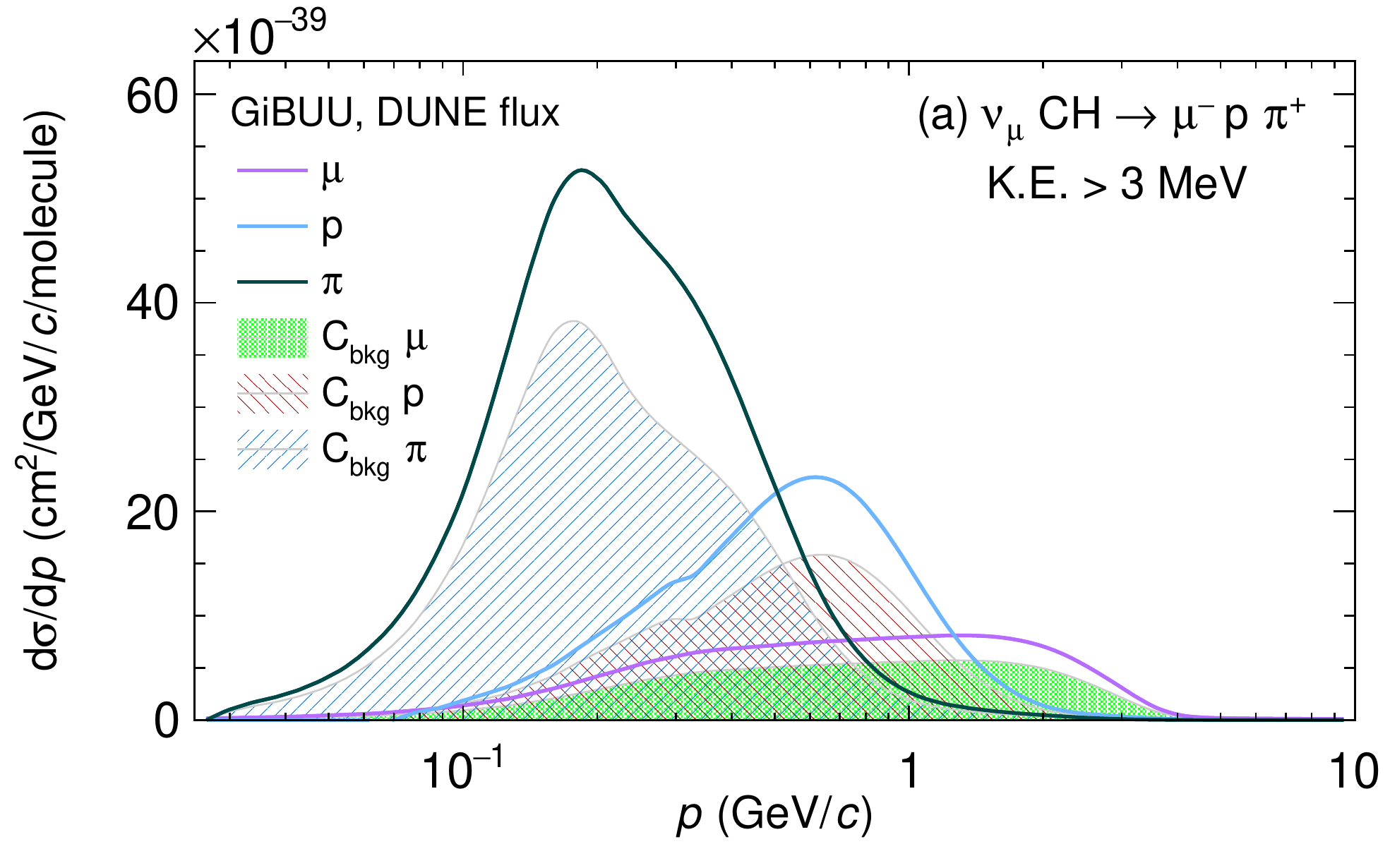}  
    \includegraphics[width=\columnwidth]{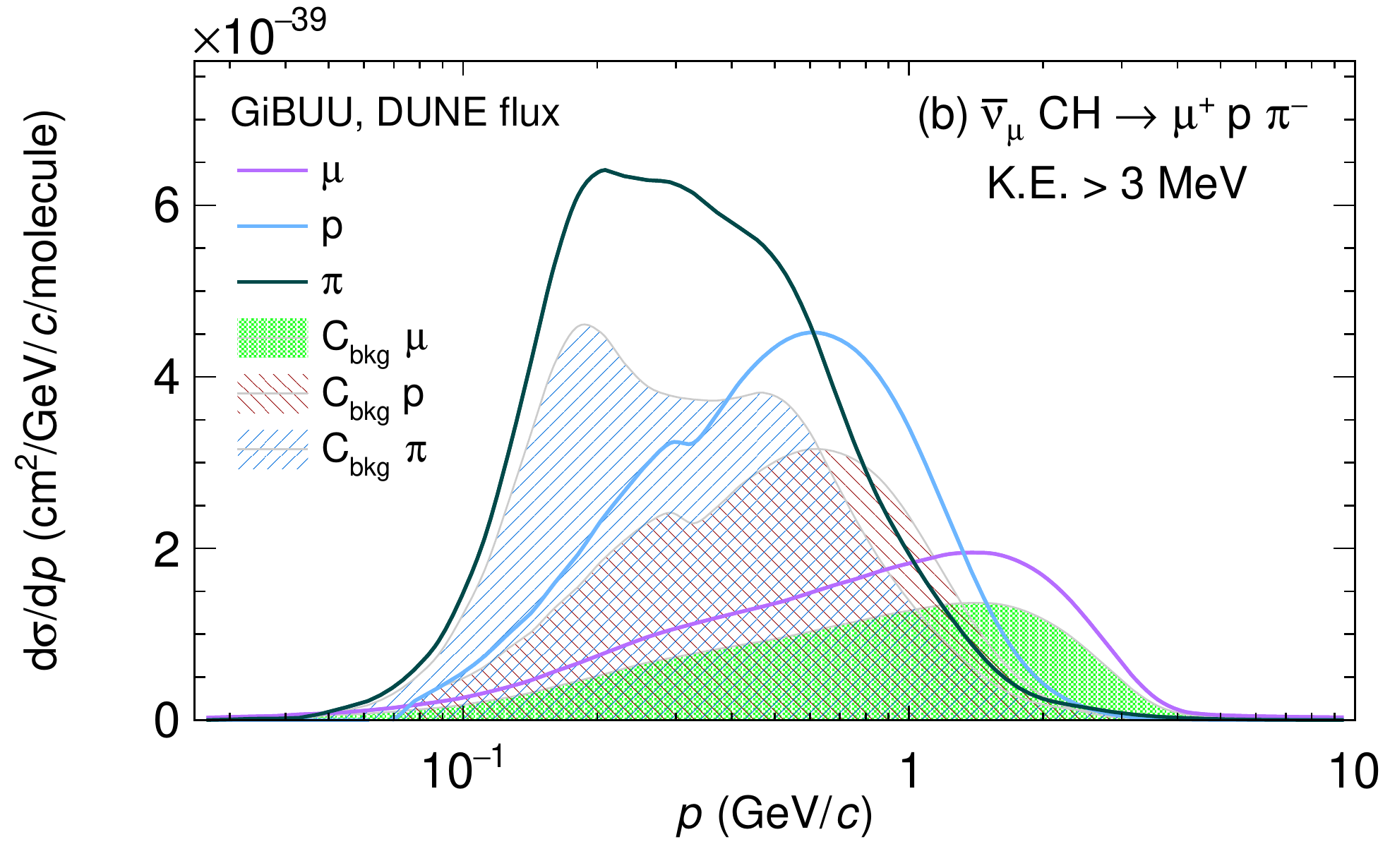}  
    \caption{Flux-averaged differential cross section as a function of the final-state particle momentum $p$ for (a) neutrino and (b) antineutrino interactions on the CH ``molecule". The pion (and the correlated proton) spectral shape in carbon results from the competition between the resonance structure and the pion absorption~\cite{PinzonGuerra:2018rju} inside the nucleus.}
    \label{fig:momentum}
\end{figure}

Without loss of generality,  consider  neutrino interactions on a  CH model ``molecule" which has the same hydrogen-carbon ratio as polystyrene [\polys],
\begin{align}
\nu+\textrm{CH}&\rightarrow\mu^-+\proton+\pi^++\textrm{X},~\textrm{and}\label{eq:nuAres}\\
\bar{\nu}+\textrm{CH}&\rightarrow\mu^++\proton+\pi^-+\textrm{X},\label{eq:nubarAres}    
\end{align}
where $\mu$, $\proton$, and $\pi$ are required to have kinetic energy greater than \unit[3]{MeV}, and $\textrm{X}$ stands for the molecular remnant. Flux-averaged differential cross sections in the polar angle $\theta$ with respect to the neutrino direction, as well as in the particle momentum $p$, are calculated using the event generator GiBUU (2019 version)~\cite{Buss:2011mx} with the DUNE fluxes~\cite{Abi:2020wmh}. As can be seen in Figures~\ref{fig:theta} and \ref{fig:momentum}, the muons are mostly at low angle and high momentum, the pions are at high angle and low momentum, and the protons, between them. As neutrinos interact with the gas  inside the TPC, high-angle events could  be detected. This is advantageous compared to the forward angular acceptance imposed by an external target to the TPC; one such example is the T2K near detector TPCs~\cite{Abe:2019arf} that measure the final-state particles from neutrino interactions on polystyrene in upstream detectors. With the full acceptance and the low  threshold, a \hptpc could detect the large majority of the final-state particles. Considering instead thresholds of \unit[100]{MeV} and \unit[75]{MeV} for protons and pions, respectively, as in a polystyrene tracker~\cite{Lu:2018stk, Mislivec:2017qfz}, only 26\% of the neutrino and 18\% of the antineutrino events would be below threshold. The higher acceptance for the antineutrino channel is due to the additional high-mass resonances [$N(1440)$, $N(1535)$ and $N(1650)$] that enhance the high-momentum parts of the spectra and therefore reduce the impact by the thresholds. 

\begin{figure}[htb]
\centering
\includegraphics[width=0.75\columnwidth]{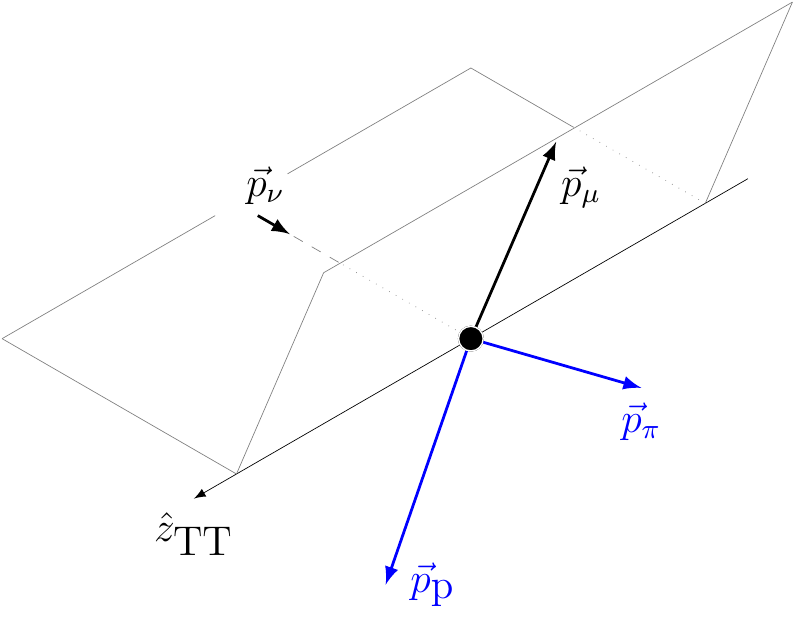} 
\caption{Schematic diagram for the particle kinematics of Eqs.~(\ref{eq:nures}) and~(\ref{eq:nubarres}). The neutrino and muon  momentum vectors, $\vec{p}_\nu$ and $\vec{p}_\mu$, define the double-transverse axis $\hat{z}_\textrm{TT}\equiv\vec{p}_\nu\times\vec{p}_\mu/|\vec{p}_\nu\times\vec{p}_\mu|$, onto which the proton and pion  momentum vectors, $\vec{p}_\proton$ and $\vec{p}_\pi$, are projected. The sum of these projections defines \dptt in Eq.~(\ref{eq:topo}). }
\label{fig:diagram}
\end{figure}

Across the whole $\theta$-$p$ phase space, the hydrogen signal and carbon background are indistinguishable. To identify the hydrogen, a three-track TKI corresponding to Eqs.~(\ref{eq:nures}) and~(\ref{eq:nubarres}), the so-called double-transverse momentum imbalance,  was introduced~\cite{Lu:2015hea}:
 \begin{align}
\dptt \equiv \left(\vec{p}_\proton + \vec{p}_\pi \right)\cdot\hat{z}_\textrm{TT} ,\label{eq:topo}
 \end{align}
  where $\hat{z}_\textrm{TT}$ is the unit vector along $\vec{p}_\nu\times\vec{p}_\mu$, and $\vec{p}_\kappa$ denotes the  momentum vector of particle $\kappa$ (Figure~\ref{fig:diagram}).

While the intrinsic \dptt on hydrogen is zero, on heavy nuclei it is dominated by Fermi motion and has a typical width of $\sim\unit[200]{\mevc}$. The measured \dptt distribution from hydrogen interactions is, therefore, a function of detector resolution, in contrast to the one from heavy nuclei, which has an irreducible width due to intranuclear dynamics. This is the basis of the TKI technique for an event-by-event selection of neutrino-hydrogen interactions with superb tracking detectors. The reconstruction resolution of \dptt by the T2K TPC is estimated to be  $\sim\unit[20]{\mevc}$~\cite{Lu:2015hea}. The T2K TPC transverse (to the magnetic field) momentum (\pt) resolution is $\oom(10\%)$ at $\pt=\unit[1]{\gevc}$~\cite{Abgrall:2010hi}.  With state-of-the-art TPC tracking performance, like that achieved with the ALICE TPC, whose \pt-resolution is $\oom(1\%)$ at \unit[1]{\gevc}~\cite{Dellacasa:2000bm, Abelev:2014ffa}, one would expect that a \dptt-resolution of $\oom(\unit[1]{\mevc})$ could be obtained. 

\begin{figure}[!htb]
    \centering
    \includegraphics[width=\columnwidth]{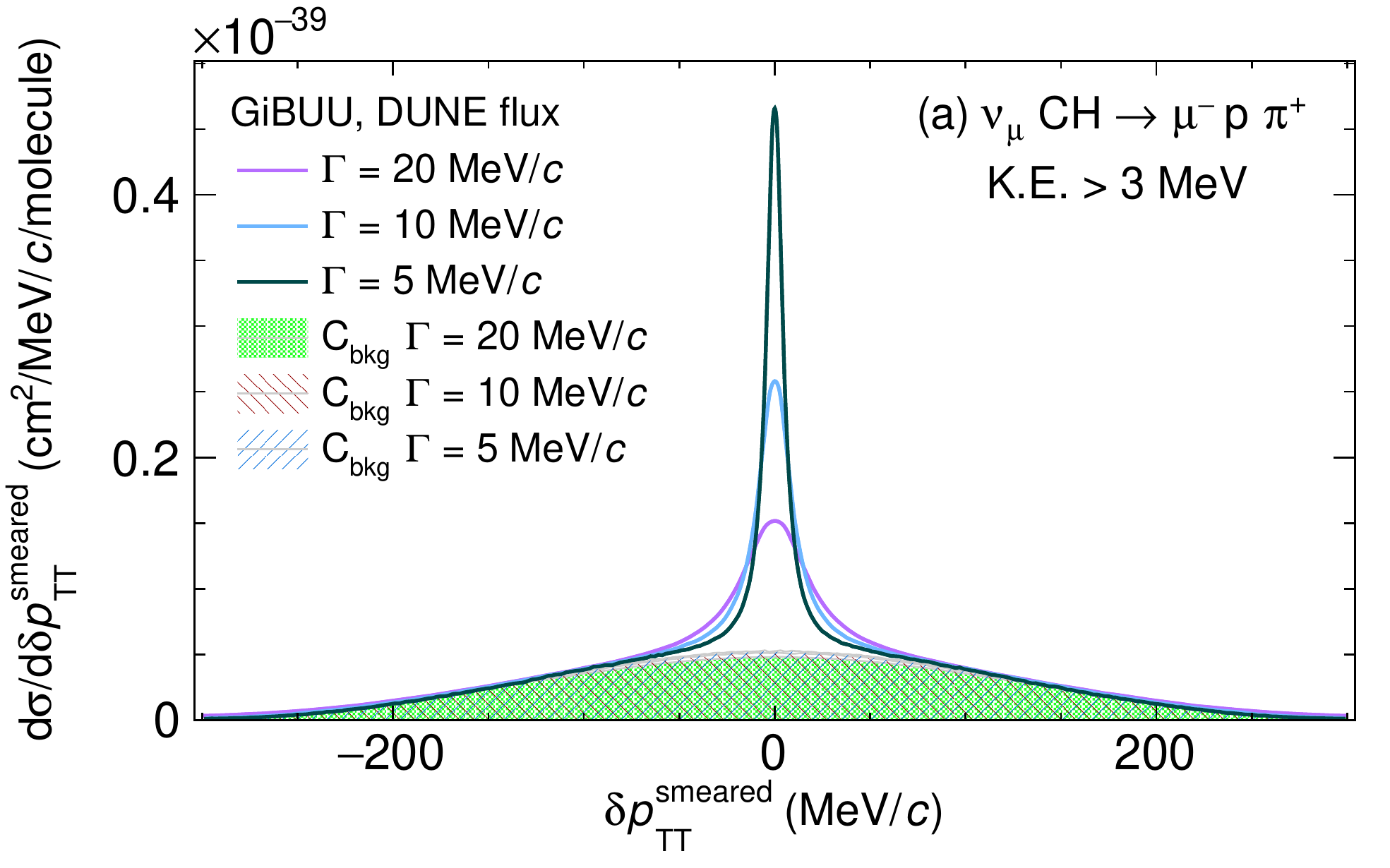}  
    \includegraphics[width=\columnwidth]{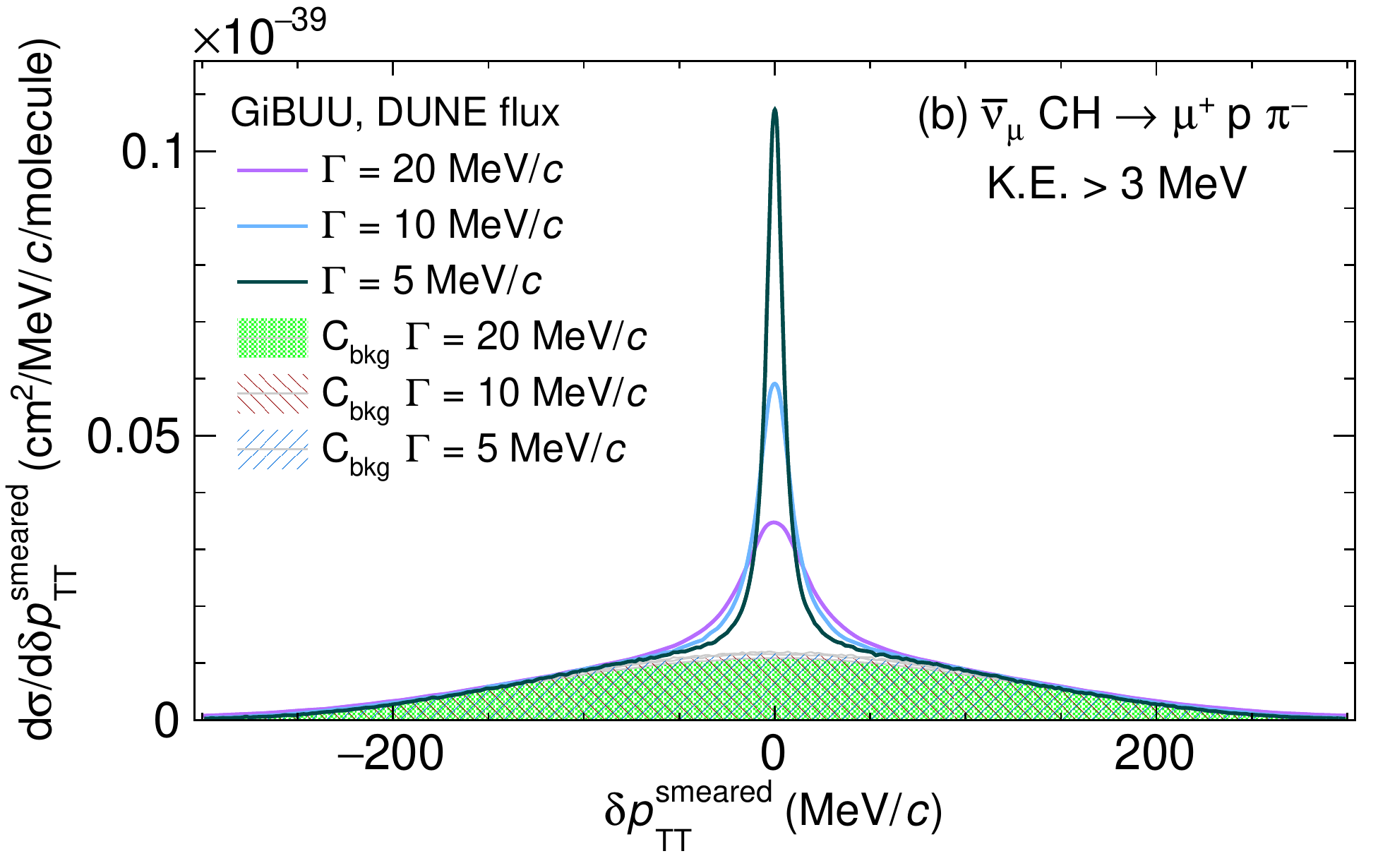}  
    \caption{Flux-averaged differential cross section as a function of the smeared \dptt for (a) neutrino and (b) antineutrino interactions on  CH.  The smearing is done by adding to the true \dptt a random variable $\epsilon$ following a Cauchy-Lorentz p.d.f. $\sim1/(\epsilon^2+\Gamma^2)$. Comparison is made among different widths, $\Gamma=\unit[5, 10,\textrm{ and }20]{\mevc}$.}
    \label{fig:nuch}
\end{figure}

To illustrate this idea, we will use the same GiBUU calculation shown above to calculate a smeared \dptt:
\begin{align}
\sdptt\equiv\dptt+\epsilon,     \label{eq:sdptt}
\end{align}
where $\epsilon$ is a random variable that follows a Cauchy-Lorentz probability density function (p.d.f.) $\sim1/(\epsilon^2+\Gamma^2)$ to mimic the effect of reconstruction resolution. The width parameter $\Gamma$ takes three values that represent different tracking performance: \unit[20]{\mevc} (as is for the T2K TPC),  \unit[10]{\mevc},  and \unit[5]{\mevc}. While, in practice, momentum resolution is commonly fit by two Gaussian functions, where the second one is needed to describe the relatively small amount of events that have large reconstruction bias, here we choose instead the Cauchy-Lorentz p.d.f.\ to provide a unified description~\cite{Lu:2015hea}. The differential cross sections in the smeared \dptt (Figure~\ref{fig:nuch}) show that, while  the hydrogen \dptt changes its Lorentzian shape with the width, the background varies insignificantly.

\begin{table}[!ht]
    \begin{ruledtabular}
        \begin{tabular}{cccccc}
            \multicolumn{2}{c}{$\left|\dptt^\textrm{smeared}\right|<3\Gamma$} & \multicolumn{2}{c}{$\sigma$ (\unit[$10^{-39}$]{cm$^2$})} &  &  \\
            & $\Gamma$ (\mevc) & $S$  & $B$ & $S/B$ & purity (\%) \\
            \hline
            \multirow{3}{*}{$\nu$ CH} & 20 &  5.4   & 5.9  & 0.92   & 48 \\
                                      & 10 &  5.4   & 3.2  & 1.7    & 63 \\
                                      & 5  &  5.4   & 1.7  & 3.2    & 76 \\
            \hline
            \multirow{3}{*}{$\bar{\nu}$ CH} & 20 &  1.2   & 1.3   & 0.93   & 48  \\
                                            & 10 &  1.2   & 0.73  & 1.7    & 63 \\
                                            &  5 &  1.2   & 0.38  & 3.3    & 77
        \end{tabular}
    \end{ruledtabular}
    \caption{Integrated cross section within $3\Gamma$ of \sdptt for   neutrino and  antineutrino interactions on  CH. For  different  $\Gamma$, the respective cross section is calculated for both the hydrogen signal $S$ and the carbon background $B$. The $S/B$-ratio and purity, $S/(S+B)$,  are also calculated.}
    \label{tab:SBCH}
\end{table}

To select the neutrino-hydrogen interactions, one could cut on $\sdptt$. To quantify the performance of such a  selection, the signal and background integrated cross section, $S$ and $B$ respectively,  within the region $\left|\dptt^\textrm{smeared}\right|<3\Gamma$ are calculated in Table~\ref{tab:SBCH}. In both neutrino and antineutrino channels, at $\Gamma=\unit[20]{\mevc}$,  the signal and background are of a similar size, yielding a $\sbr$-ratio about 1. At a four-fold  reduction of $\Gamma$, the calculated $\sbr$-ratio  reaches 3.2, the corresponding purity [$S/(S+B)$] being 76\%. On the one hand, it is important  to point out that these numbers depend on the modeled nuclear effects. More generally speaking, the departure of the hydrogen-carbon cross-section ratio from $1/6$ is a measure of the nuclear medium  effects~\cite{Lu:2015hea}.   As the $\sbr$-ratio is affected by FSI on top of the Fermi motion of the initial bound proton, mismodeled FSI such as the elastic component of GENIE \textit{hA}~\cite{Lu:2015tcr, Abe:2018pwo, Lu:2018stk, Harewood:2019rzy, Cai:2019hpx} could cause significant bias~\cite{Pickering:2016mlu}. In the current GiBUU calculation, because  $\pi^+$ and $\pi^-$ experience  very similar FSI   inside the carbon remnant, even though both the signal and background size are different between the neutrino and antineutrino channels, the $\sbr$-ratio is shown to be very similar  between the two. 
On the other hand, regardless of the underlying nuclear effects,  for the same size of signal, the background size decreases with $\Gamma$. As the relative size of the background is reduced---via an improvement of the tracking resolution and an increase in the hydrogen content as discussed in the following sections---the relative background uncertainty  will decrease  and become insignificant. 

\section{Gas mixture candidates}\label{sec:method}

\subsection{ Argon-Alkane Mixtures}\label{sec:aral}

\begin{figure}[tb]
    \centering
    \includegraphics[width=\columnwidth]{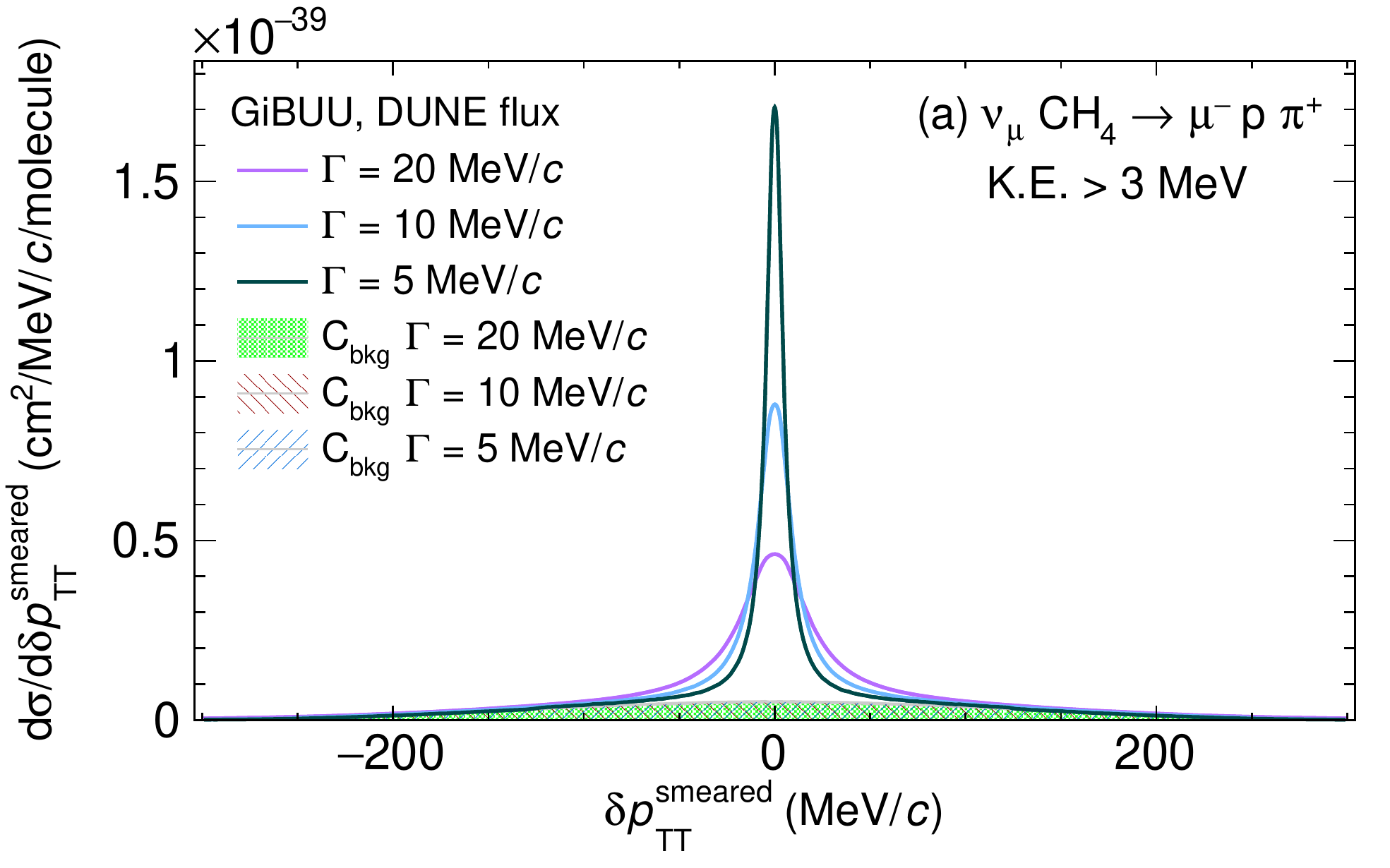}  
    \includegraphics[width=\columnwidth]{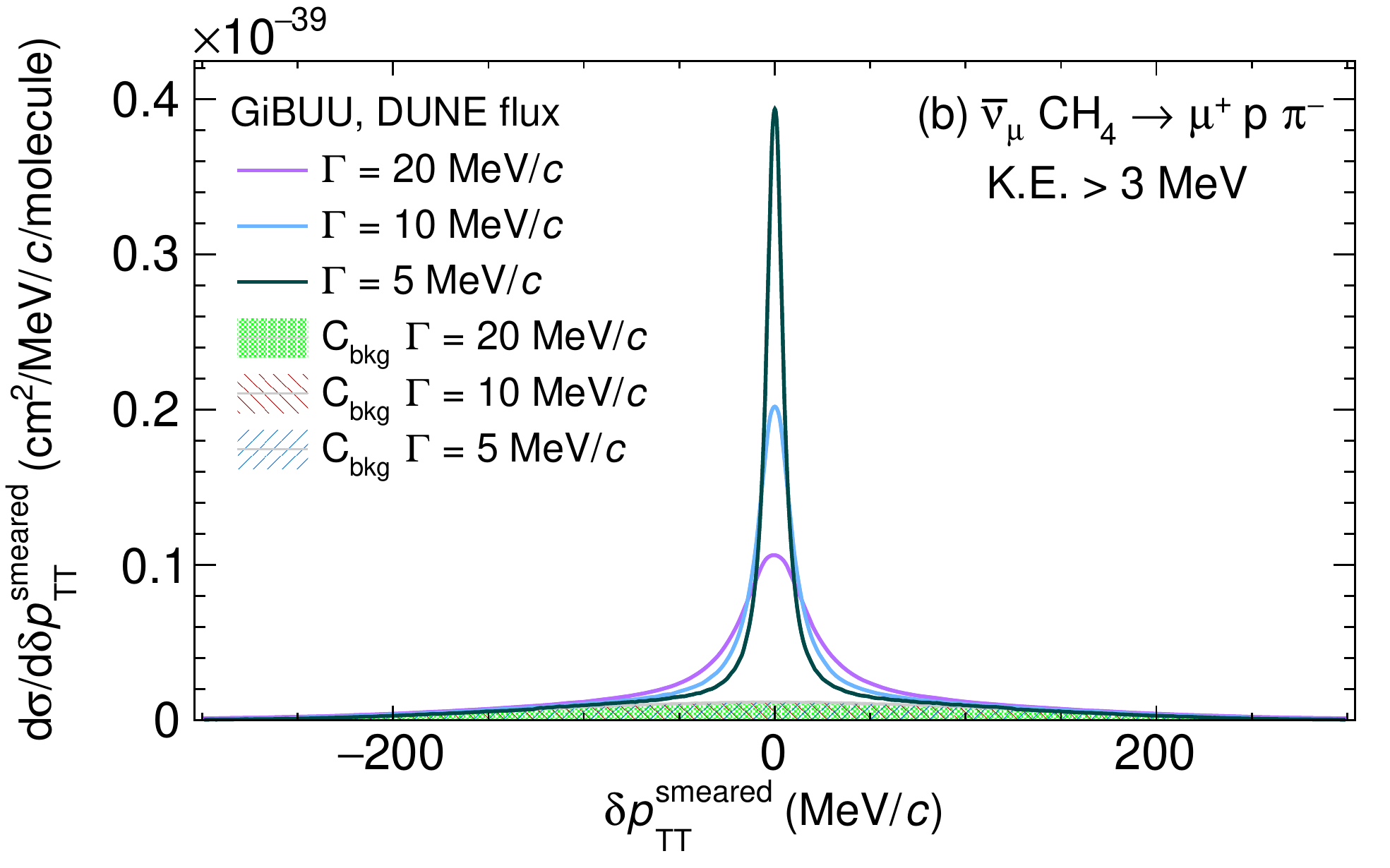}  
    \caption{Flux-averaged differential cross section as a function of the smeared \dptt for (a) neutrino and (b) antineutrino interactions on \methane.}
    \label{fig:nuch4}
\end{figure}

Gas mixtures for TPCs have long been studied  in field regions suitable for drift and gas amplification (cf. Refs.~\cite{Peisert:1984xj, Rolandi:2008qla, Hilke:2010zz}). Their typical composition is a noble gas with one or more admixtures of other gases to engineer drift properties, like drift velocity and diffusion, for the intended detector geometry and event characteristics. Organic molecules like alkane in the admixture stabilize the gas amplification by suppressing UV photons in the avalanche gas amplification process, hence the name quenchers. In the case of Ar-alkane mixtures, the quencher reduces diffusion and can increase the drift velocity (see Section~\ref{sec:driftgas} for detail). Quenchers alone can also act as counter gases in TPCs. This turns out to be advantageous for the measurement of neutrino-hydrogen interactions. For example, with a pure \methane target in comparison to  CH, the hydrogen mass is increased by a factor of four for the same amount of carbon background. The calculated differential  and integrated  cross sections  are shown in Figure~\ref{fig:nuch4} and Table~\ref{tab:SBCH4}, respectively.  An $\sbr$-ratio of 13 and  a selection purity of 93\% are achieved thanks to the four-fold increase in the signal size.

\begin{table}[!t]
    \begin{ruledtabular}
        \begin{tabular}{cccccc}
            \multicolumn{2}{c}{$\Gamma=\unit[5]{\mevc}$}  & \multicolumn{2}{c}{$\sigma$ (\unit[$10^{-39}$]{cm$^2$})} & & \\
            \multicolumn{2}{c}{$\left|\dptt^\textrm{smeared}\right|<3\Gamma$}  & $S$  & $B$ & $S/B$ & purity (\%) \\
            \hline
            \multirow{2}{*}{$\nu$} & CH & 5.4   & 1.7  & 3.2  & 76  \\
                             & \methane & 22    & 1.7  & 13   & 93  \\
            \hline
            \multirow{2}{*}{$\bar{\nu}$} & CH & 1.2  & 0.38  & 3.3  & 77  \\
                                   & \methane & 5.0  & 0.38  & 13   & 93 
        \end{tabular}
    \end{ruledtabular}
    \caption{Integrated cross section within $3\Gamma$ ($\Gamma=\unit[5]{\mevc}$) of the smeared \dptt for   neutrino and  antineutrino interactions on CH  and \methane.    }
    \label{tab:SBCH4}
\end{table}

While \ppten has been a common choice as TPC gas (see, for example, Ref.~\cite{Wieman:1997rs}),  other gases have also been used. For example, the ALICE TPC uses \alicegas to cope with the high multiplicity  environment at very high event rates in heavy-ion collisions~\cite{Dellacasa:2000bm}, while the T2K near detector TPCs use \ttokgas~\cite{Abgrall:2010hi}. Both examples have been operating at atmospheric pressure. 

\begin{figure}[!ht]
    \centering
    \includegraphics[width=\columnwidth]{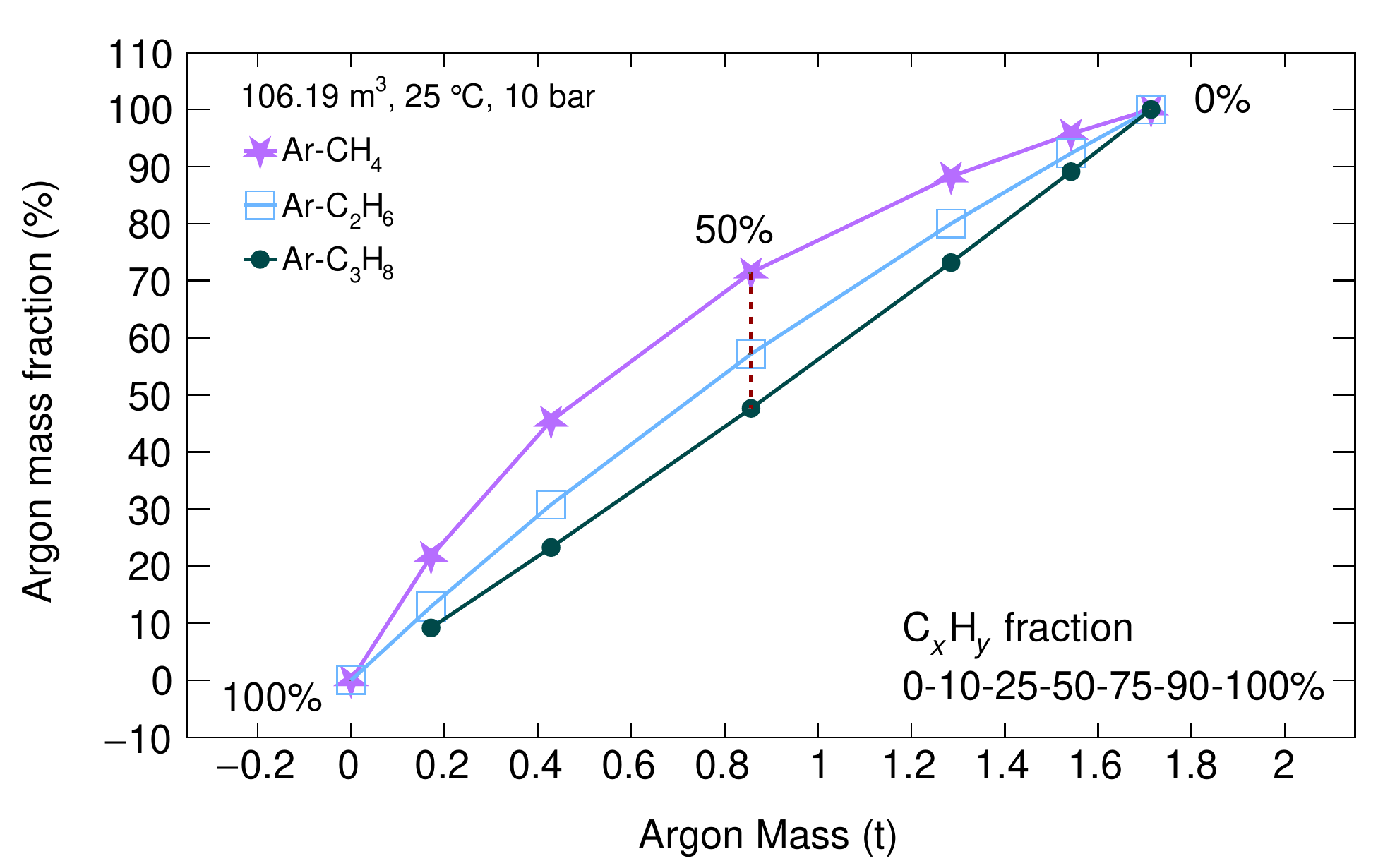} 
    \caption{Argon mass fraction vs. argon mass for different Ar-alkane mixtures at \unit[25]{$^\circ$C} and \unit[10]{bar} in a volume of \unit[106.19]{$\textrm{m}^3$} [$\pi\cdot(\unit[2.6]{m})^2\times\unit[5]{m}$]. The different points along respective solid lines are a scan of the alkane ($\cphq$) concentration: 0\%, 10\%, 25\%, 50\%, 75\%, 90\%, and 100\% (Ar-\propane only up to 90\%, see text). Different alkane points at the same fraction (50\%) are connected by a dashed line.}
    \label{fig:argon_content}
\end{figure}

In DUNE, in order to provide constraints on neutrino interactions on argon in the far detectors, \ppten is the default gas mixture of the near detector \hptpc. Therefore, we focus on argon-based, in particular Ar-alkane, mixtures as an extrapolation of the default gas. Depending on the  argon mass fraction (that is, the argon purity in terms of mass), one could choose a certain argon concentration  for the desired argon mass. For example, as is shown in Figure~\ref{fig:argon_content}, both \ppft  (\pft) and \propft have the same argon mass, but the argon mass fraction in \ppft is higher by a relative 50\%. 

In the following, we discuss  a range of Ar-alkane mixtures as \hptpc gas candidates   by examining their  hydrogen content and tracking-related  properties. Unless otherwise specified,  we fix the temperature at \unit[25]{$^\circ$C}  throughout the discussions.

\subsection{Hydrogen content}\label{sec:hydrogen}
 
In the previous calculation for a CH target,  a $\unit[5]{\mevc}$ \dptt-resolution leads to a selection $\sbr$-ratio of 3.2, corresponding to a purity of 76\%.  In a \hptpc, as one can choose a variety of gas mixtures, the selection purity, which depends on the  ratio between the number of free protons and that of the bound ones, can be optimized alongside with the hydrogen mass as follows.
 
\begin{figure}[!t]
    \centering
    \includegraphics[width=\columnwidth]{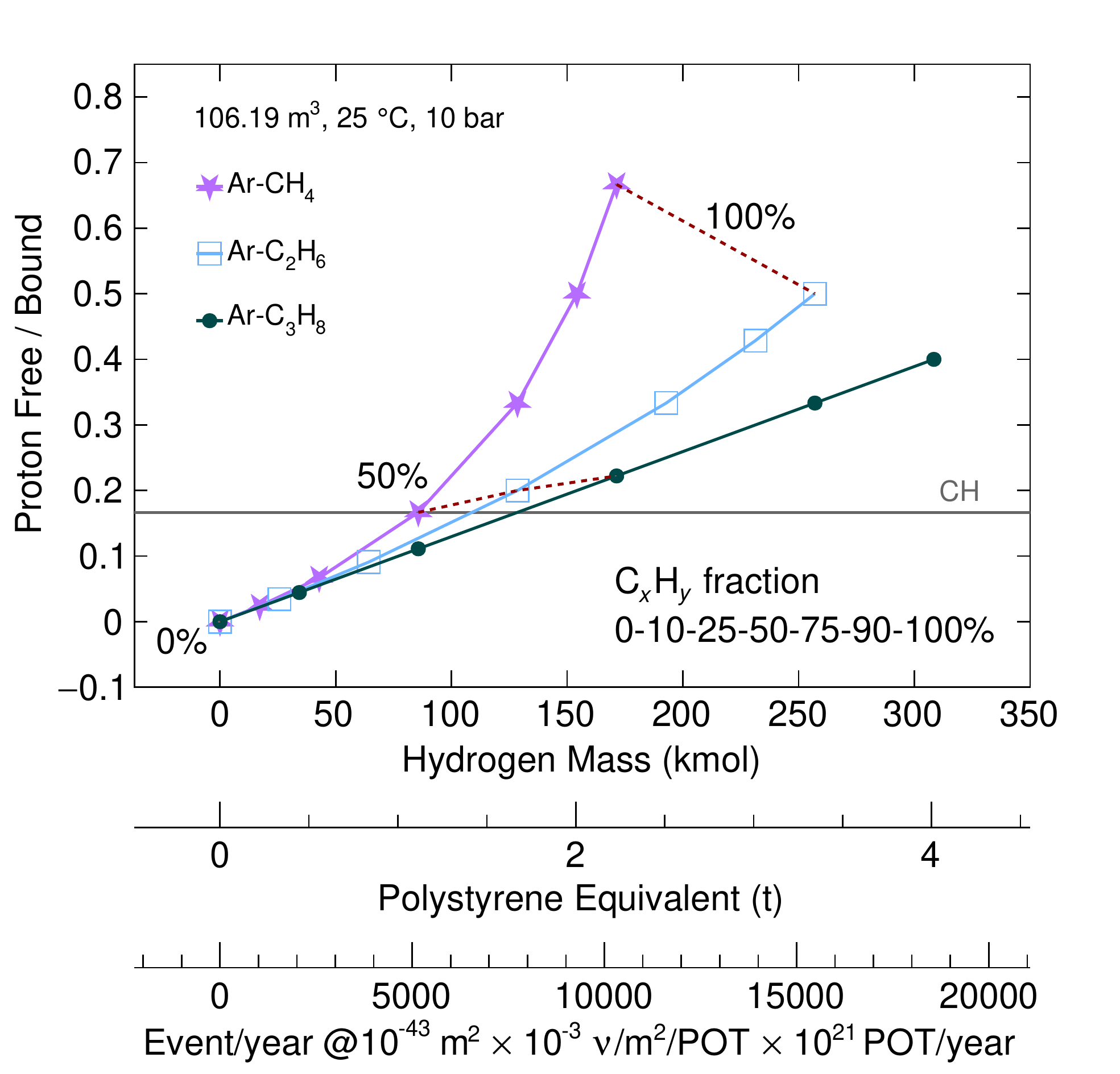} 
    \caption{Proton free-to-bound  ratio vs.\ hydrogen mass for different Ar-alkane mixtures. As a comparison, the ratio for polystyrene (CH) is indicated by a horizontal line regardless of the hydrogen mass.  The hydrogen mass is converted to the hydrogen-equivalent polystyrene mass, as well as to  the neutrino-hydrogen event rate per year, assuming a cross section of $\unit[10^{-43}]{\textrm{m}^2}$ which is typical for the signal channels [Eqs.~(\ref{eq:nures}) and~(\ref{eq:nubarres})], and an exposure of $10^{21}$  protons on target (POT) per year with a  flux of  $\unit[10^{-3}]{\textrm{neutrinos}/\textrm{m}^2/\textrm{POT}}$ that gives the projected neutrino rate at DUNE~\cite{Abi:2020wmh}.}
    \label{fig:hydrogen}
\end{figure}

A CH target has a proton free-to-bound  ratio of $1/6$,  the same as in \ppft (Figure~\ref{fig:hydrogen}). This ratio increases with the methane concentration and reaches $1/2$ for \ppnty (\pnty), and $2/3$ for pure \methane. By replacing  CH (or \ppft) with pure \methane as the interaction target, the $\sbr$  ratio is shown to be  improved by a factor of $(\nicefrac{2}{3})/(\nicefrac{1}{6}) = 4$ (Table~\ref{tab:SBCH4}).

One mole of \ppft  has the same amount of hydrogen  as one mole of $\textrm{H}_2$. In addition, 
as is shown in Figure~\ref{fig:hydrogen} at \unit[10]{bar} in a volume of \unit[106.19]{m$^{3}$}, the \ppft gas contains  the same hydrogen mass as  \unit[$\sim1$]{t} of polystyrene. This  amounts to $\sim10\%$ of the proposed DUNE 3DST detector  that is a polystyrene tracker with dimensions  $\unit[2.4\times2.4\times2]{\textrm{m}^3}$~\cite{Abi:2020wmh}. With the projected DUNE beam exposure, $10^{21}$ protons on target (POT) per year and  $\unit[10^{-3}]{\textrm{neutrinos}/\textrm{m}^2/\textrm{POT}}$~\cite{Abi:2020wmh}, this hydrogen mass gives $\sim5000$ three-track events [Eqs.~(\ref{eq:nures}) and~(\ref{eq:nubarres})] per year  assuming a  typical cross section of  $10^{-43}~\textrm{m}^2$ (Table~\ref{tab:SBCH}, cf. also Ref.~\cite{Tanabashi:2018oca}). It follows immediately that pure \methane improves the hydrogen event rate by a factor of 2, to $\sim10^4$ per year. 

As alkanes (\cphq, $y=2x+2$) are acyclic saturated hydrocarbons, it follows that \methane with $x=1$ provides the highest proton free-to-bound ratio among all hydrocarbons. 
For a given concentration, other members in the alkane series like ethane (\ethane) and propane (\propane) can provide larger  hydrogen mass with a different proton free-to-bound ratio (Figure~\ref{fig:hydrogen}). For example, \propft has twice the hydrogen mass as \ppft. However, this progress along the series  is limited by the phase boundaries of the gas candidates. At \unit[25]{$^\circ$C} and \unit[10]{bar}, the maximal concentration of \propane is  \unit[\maxpropane]{\%} and for isobutane (\isobutane) it is \unit[\maxisobutane]{\%}~\cite{phasediag}---higher than these they liquefy.  Therefore, \propnt provides the maximal hydrogen mass among all Ar-alkane candidates.

\section{Tracking-related gas properties}\label{sec:driftgas}

The measurement of neutrino-hydrogen interactions---Eqs.~(\ref{eq:nures}) and~(\ref{eq:nubarres})---from an Ar-alkane gas mixture via the TKI technique relies on the reconstruction of the trajectories in the TPC of primary charged particles ($\mu^\mp$, $\proton$, and $\pi^\pm$). When these particles traverse the detector volume, they ionize the gas liberating electrons that are then driven towards the readout plane under the influence of electric and magnetic fields. The arrival time, position, and amplitude of the collected drift-electron signals are used to reconstruct the trajectories and characteristic energy loss (\dEdx) of the primary particles. In the presence of a magnetic field, the particles' charges and momenta can also be measured.

During their propagation, the drift electrons collide with the gas molecules at energy and time scales different from those of the primary ionization.
The rate of these collisions depends on the gas density, which is sensitive to temperature (\temp) and pressure (\pres), and thus running detectors are regularly calibrated towards a certain operational point via temperature and pressure scaling (see, for example, Ref.~\cite{Abgrall:2010hi}). Such density corrections for the drift field ($E$) and the gas parameters that we will discuss in this Section are given in Table~\ref{tab:density_corrections}~\cite{Peisert:1984xj,Rolandi:2008qla,Gonzalez-Diaz:2017gxo}. 

\begin{table}[htb]
    \begin{ruledtabular}
        \begin{tabular}{lr}
            \textrm{Drift field and gas parameters}          & \textrm{Density correction} \\
            \hline
            Electric field strength & \ETp \\
            Drift velocity                  & \vd  \\
            Diffusion coefficients          & $\dlt\cdot\sqrt{\nicefrac{\pres}{\temp}}$   \\
            First Townsend coefficient  & $\alpha\cdot\nicefrac{\temp}{\pres}$ \\ 
        \end{tabular}
    \end{ruledtabular}
    \caption{
        Density corrections~\cite{Peisert:1984xj,Rolandi:2008qla, Gonzalez-Diaz:2017gxo} for the electric field and the gas  parameters that  will be discussed in this Section. These scaling laws indicate that, at the same $\ETp$, the drift velocity is independent of \temp and \pres, while the diffusion decreases  and the gas gain  increases as $\nicefrac{\pres}{\temp}$ increases. 
    }
    \label{tab:density_corrections}
\end{table}

\begin{figure}[tbh]
    \centering
    \includegraphics[width=\columnwidth]{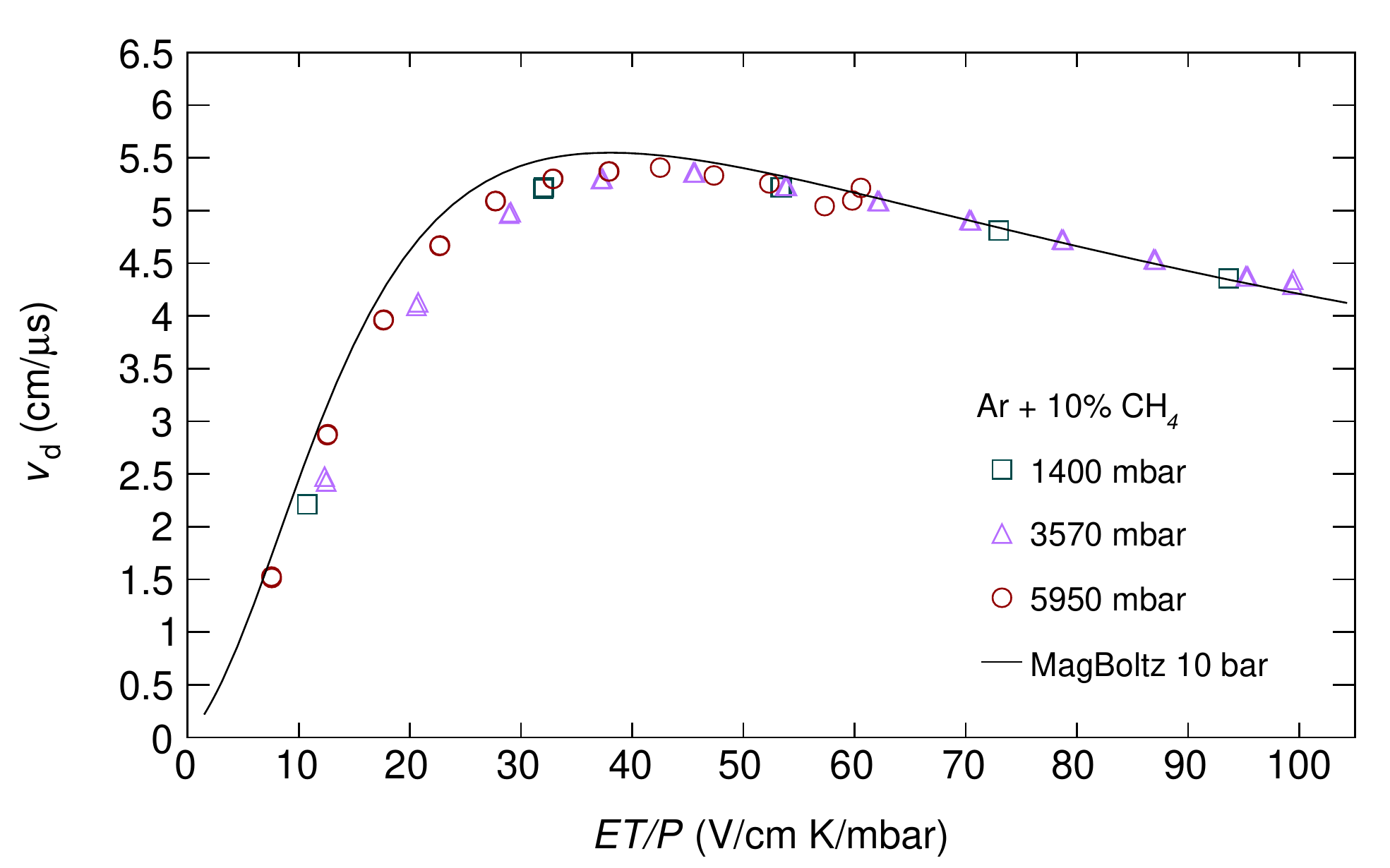}  
    \caption{Measured drift velocity  of \ppten at three pressures, from close to atmospheric up to $\sim\unit[6]{bar}$ at \unit[296--297]{K}~\cite{hamacherbaumann:2020hpg}.  The data closely follow the pressure scaling law (Table~\ref{tab:density_corrections}) over a span of \unit[5]{bar}. For comparison, the calculated drift velocity at \unit[10]{bar} by \magboltz~\cite{Biagi:1999nwa} (interfaced to \garfield~\cite{schindler:2020}) is also shown. The error bars are statistical only and smaller than the marker size.}
    \label{fig:hpgmc_p10_measurement}
\end{figure}

The discussion of   the  Ar-alkane  properties, including drift and gas gain, will be focused on their  impact on TPC performance. All calculations presented here were performed using \magboltz version 11.7~\cite{Biagi:1999nwa} interfaced to \garfield~\cite{schindler:2020}. While the temperature was set to \unit[298]{K}, results with various electric field strengths at \unit[1]{bar} or \unit[10]{bar} are compared. The \cphq fraction in the Ar-alkane mixture is scanned from \unit[0]{\%} to \unit[100]{\%}, with the exception of propane (\propane) that liquefies above \unit[\maxpropane]{\%} at \unit[10]{bar}.  The effect of a  magnetic field parallel to the electric field is explicitly discussed only when relevant.    

As a validation, the calculated drift velocity (more detail in Section~\ref{ssec:vd}) for \ppten is compared to the measurements by the  High Pressure Gas Monitoring Chamber~\cite{hamacherbaumann:2020hpg}.
Figure~\ref{fig:hpgmc_p10_measurement} shows the experimental data  in three pressure settings up to $\sim\unit[6]{bar}$. After correcting for the temperature and pressure, the data show the expected scaling behavior over the full measurement range. The \magboltz calculation reproduced the measurements satisfactorily, except for fields below $\unit[40]{\uETp}$, where the predicted drift velocity is higher by $\sim\unit[5]{\%}$; a similar deviation has been reported in Ref.~\cite{Atoum:2019jkl}.

\subsection{Drift Velocity} \label{ssec:vd}

\begin{figure}[!b]
    \centering
    \includegraphics[width=\columnwidth]{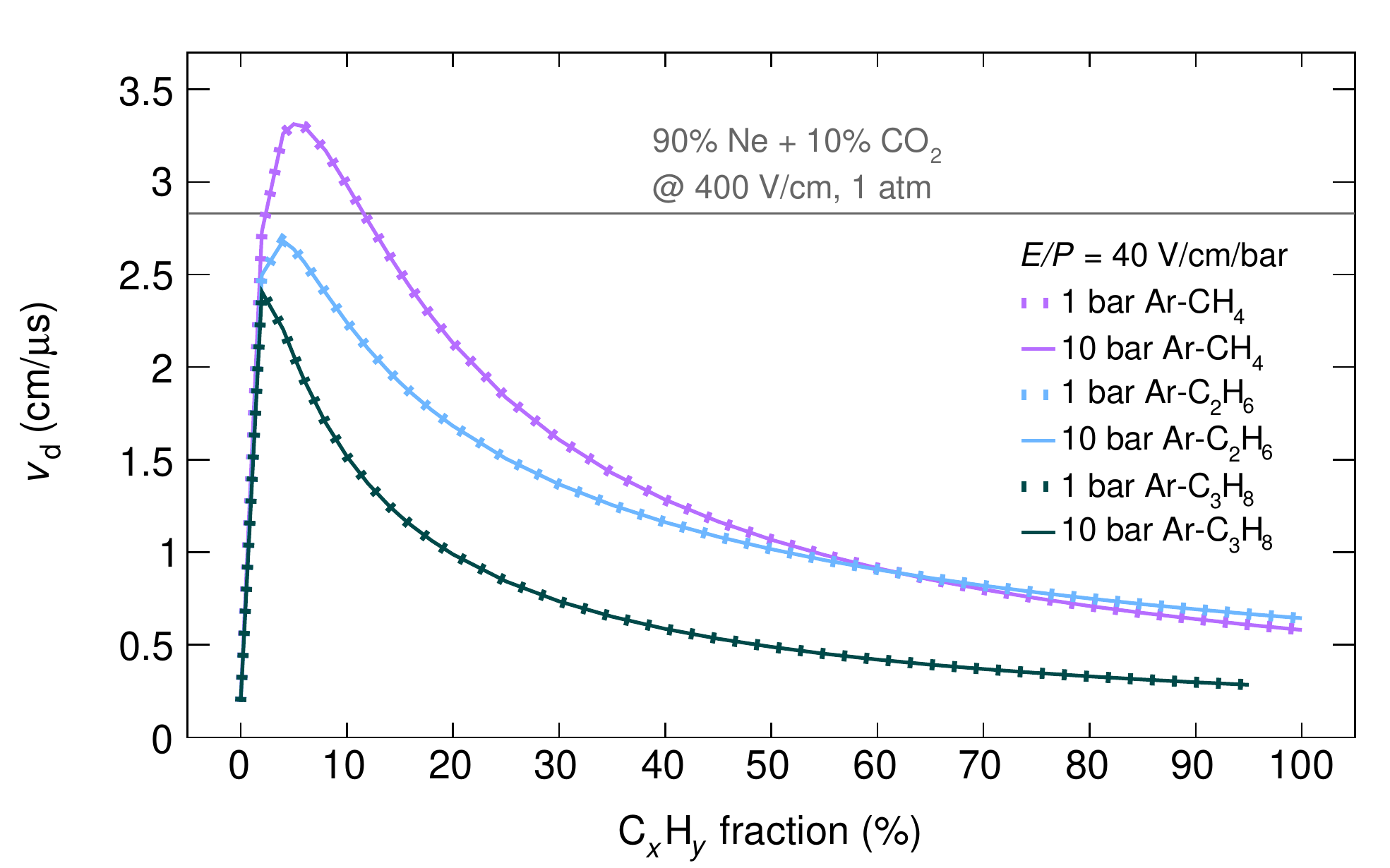}   
    \caption{ \magboltz~\cite{Biagi:1999nwa} (interfaced to \garfield~\cite{schindler:2020}) calculation of the drift velocity   at \unit[40]{\uEp} as a function of the alkane concentration.  Ar-alkane mixtures  at \unit[1]{bar} and \unit[10]{bar} are compared. The ALICE (\alicegas) value~\cite{Dellacasa:2000bm} is indicated by a horizontal line. Note the \Ep-scaling  (see also Table~\ref{tab:density_corrections}).}
    \label{fig:vd_cphq_400vcm}
\end{figure}

In a TPC, the electron drift velocity (\vd) is used to convert the signal arrival time to a position along the drift direction, enabling the three-dimensional reconstruction of the primary ionization spatial coordinates. The ALICE TPC has a drift velocity of \unit[2.83]{\uDrift} for an electric field strength $E=\unit[400]{V/cm}$ across  a drift length of \unit[2.5]{m} and for a gas mixture of \alicegas at one atmospheric pressure \cite{Dellacasa:2000bm}. Due to the \Ep-scaling (Table~\ref{tab:density_corrections}), to maintain the same drift velocity in the same gas at \unit[10]{bar}, a field strength of $\unit[4]{kV/cm}$ is required, implying a cathode voltage of \unit[1]{MV} across a 2.5-m drift length. However, as commercial  power supplies are not readily available above \unit[500]{kV}, we consider $\Ep\sim\unit[40\textrm{--}200]{\uEp}$ a practical operational region in an ALICE-sized TPC at \unit[10]{bar}. In comparison, the T2K TPC has a drift velocity of \unit[7.8]{\uDrift} at \unit[275]{V/cm} for a mixture of \ttokgas at atmospheric pressure \cite{Abgrall:2010hi}.

\begin{figure}[tb]
    \centering
    \includegraphics[width=\columnwidth]{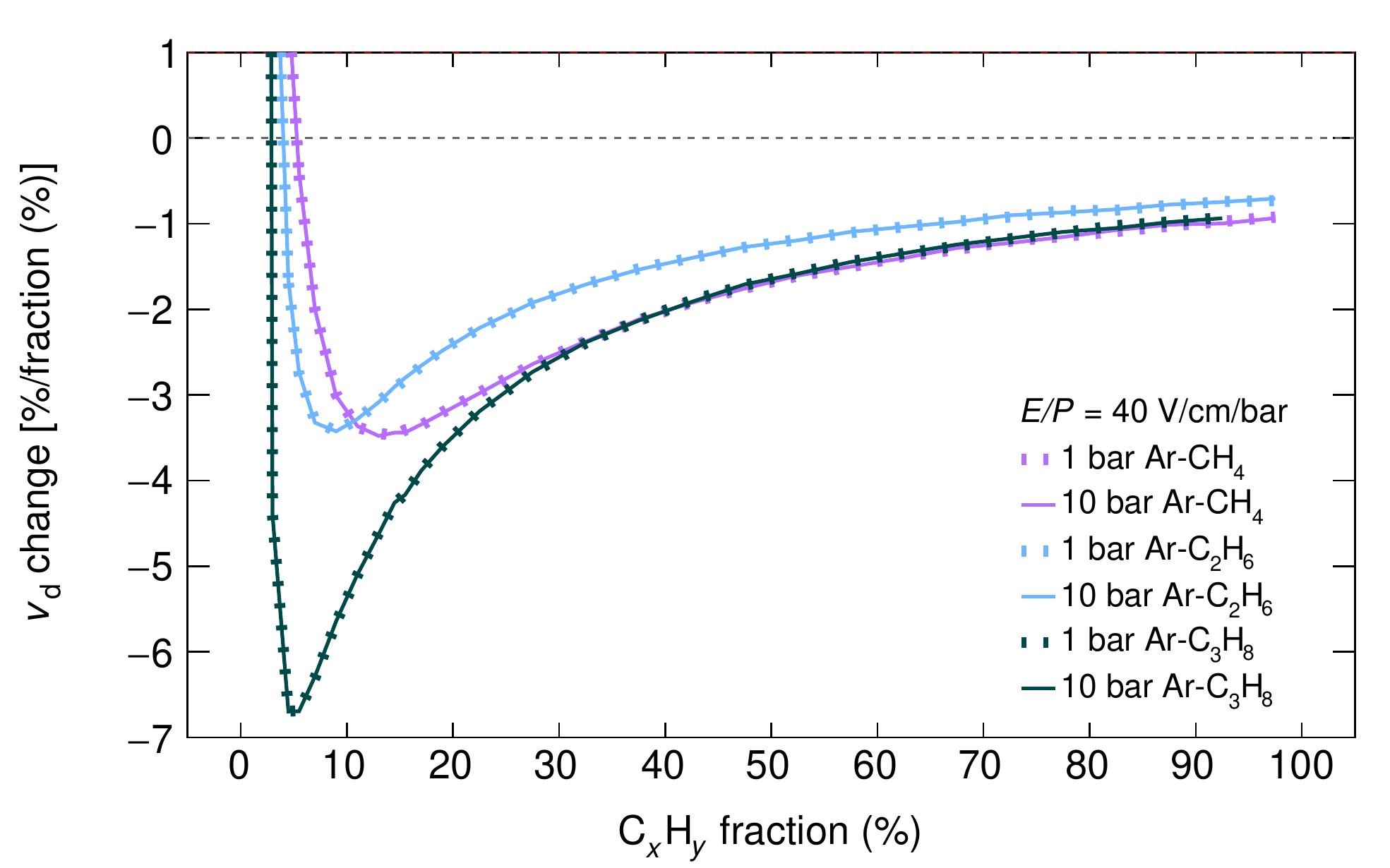}   
    \caption{Fractional change (in \%) of drift velocity  for every  percentage increase of the alkane concentration.}
    \label{fig:vd_change}
\end{figure}

\begin{figure}[tb]
    \centering
    \includegraphics[width=\columnwidth]{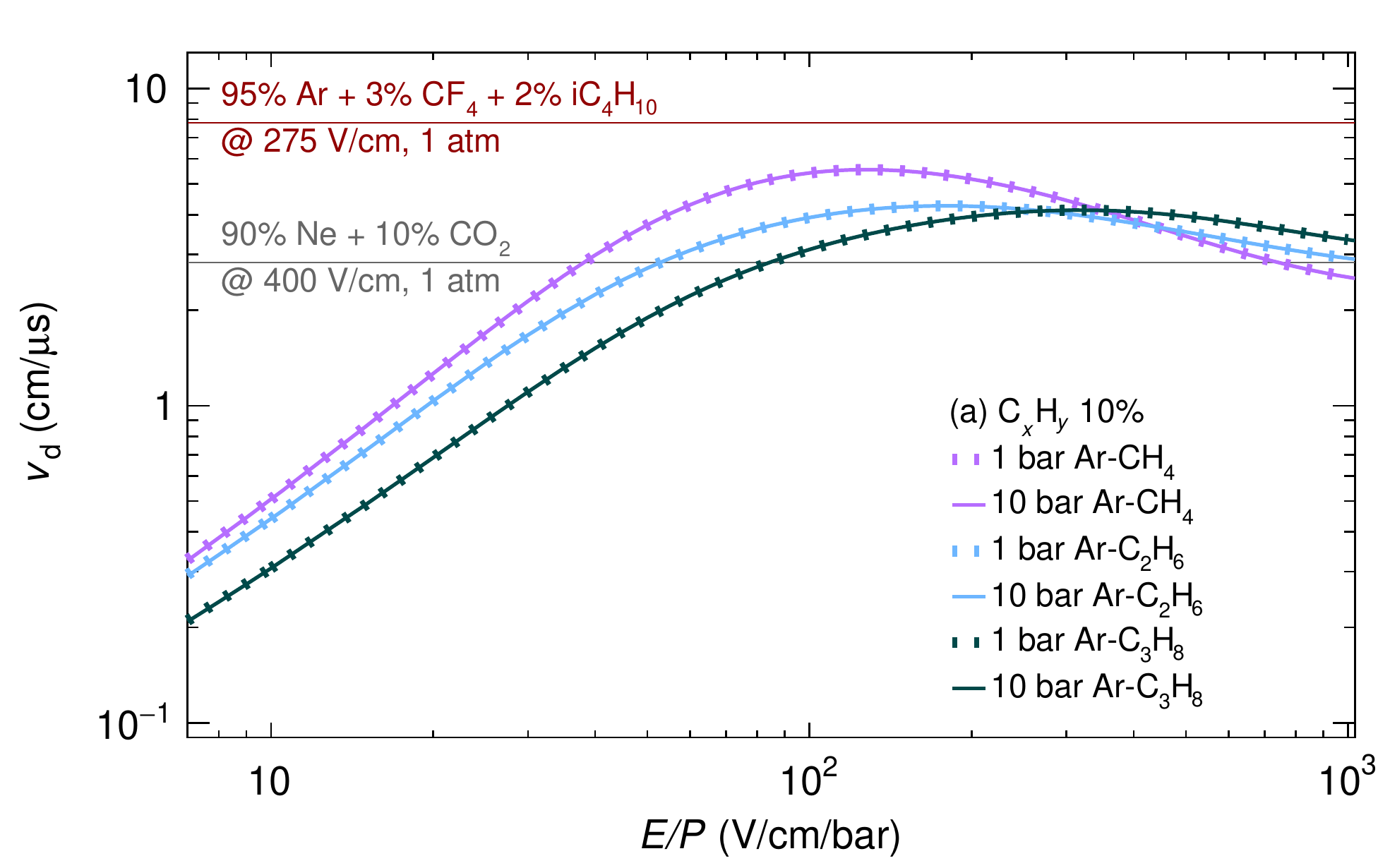} 
    \includegraphics[width=\columnwidth]{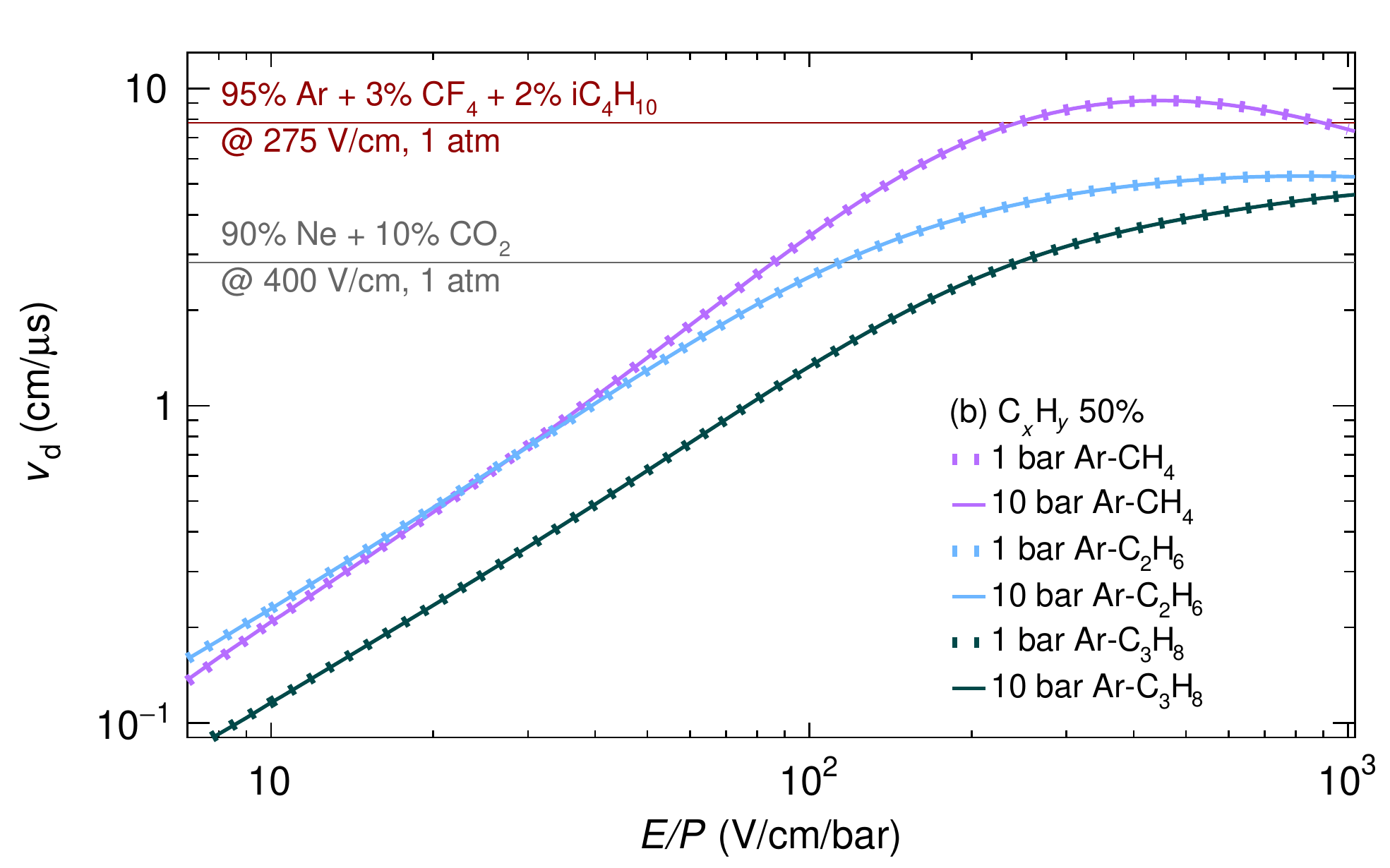} 
    \includegraphics[width=\columnwidth]{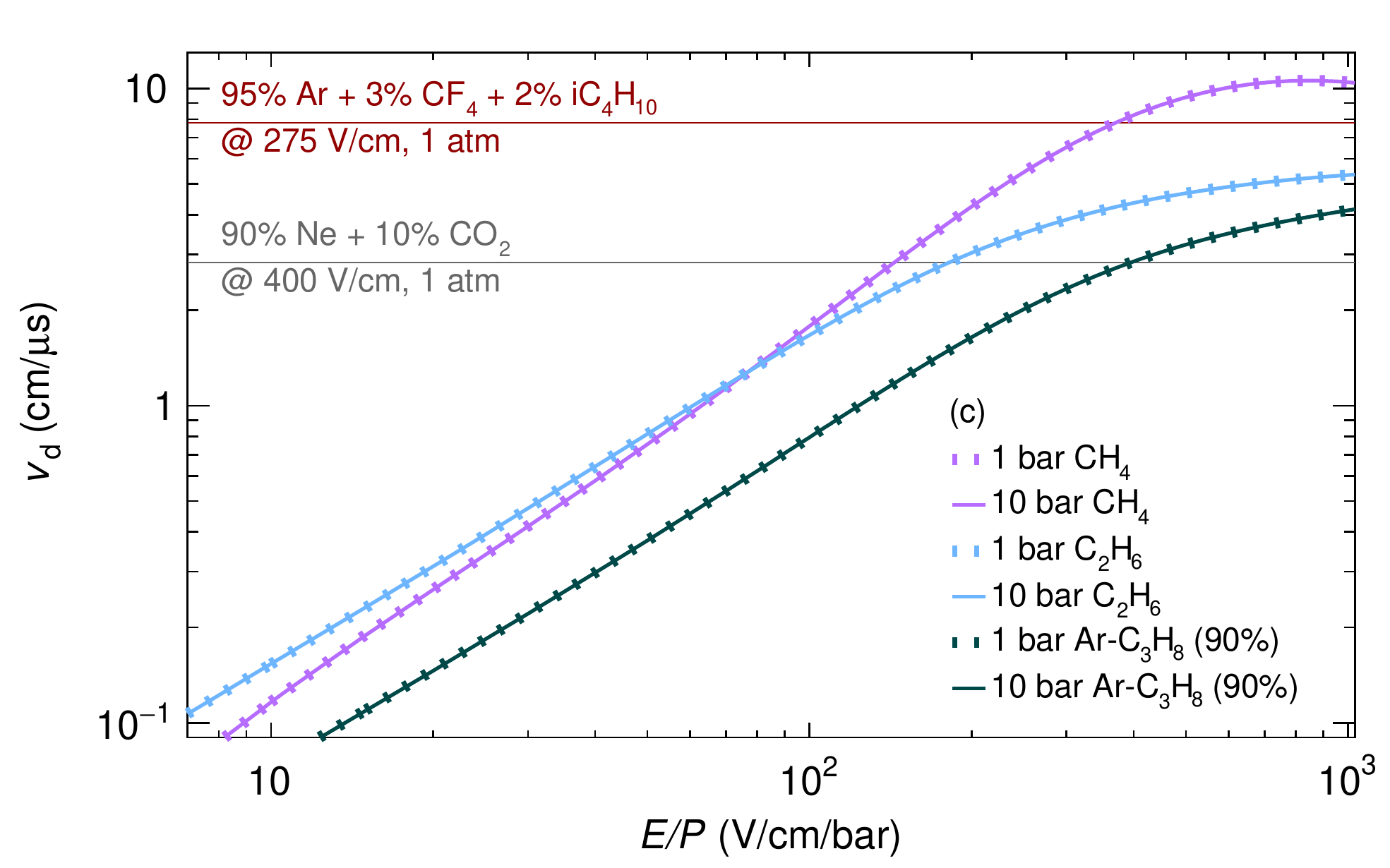} 
    \caption{Drift velocity as a function of the scaled electric field strength for different gas mixtures with  alkane concentrations  (a) 10\%, (b) 50\%, and (c) 100\% (90\% for \propane). The reference values  from ALICE and T2K (\ttokgas)~\cite{Abgrall:2010hi} are shown.}
    \label{fig:vd_Ep}
\end{figure}%

The calculated drift velocity at \unit[40]{\uEp} is shown in Figure~\ref{fig:vd_cphq_400vcm}. At a few percentage of alkane concentration, the drift velocity dramatically increases from the pure-argon value \unit[0.2]{\uDrift} by an order of magnitude. This is due to the so-called Ramsauer minimum of argon~\cite{Ramsauer:1921}: the low excitation energy of alkanes (due to the presence of the vibrational and rotational degrees of freedom) effectively reduces the energy of the drifting electrons such that the collisional  cross section of  electrons on argon reaches a minimum,  making the gas as a whole more transparent to the drifting electrons.
The drift velocity then falls back as the concentration increases, approaching \unit[0.5]{\uDrift} for pure methane and ethane, and even lower for propane-rich mixtures (still higher than for pure argon). This level of drift velocity corresponds to sub-millisecond drift time across a \unit[2.5]{m} drift length, which would allow for a pile-up-free event rate of $\mathcal{O}(\unit[1]{kHz})$, much higher than the ones foreseen in future accelerator neutrino experiments~\cite{Abi:2020wmh}.

While the magnitude of the drift velocity is not critical here, an optimal tracking performance relies on  a uniform and stable drift velocity in the large gas volume, which in turn poses a constraint on the gas  system~\cite{Kochenda:2002zz}. Figure~\ref{fig:vd_change} shows the fractional change of the drift velocity for every percentage increase of the alkane concentration. At \unit[40]{\uEp} for any \cphq concentration above \unit[5]{\%}, the change of the drift velocity is greater than \unit[1]{\%}, which indicates that a per-mil-level stability of the drift velocity  requires a control on the quencher concentration at the per-mil level.  

In the practical  operational region \unit[40--200]{\uEp}, the drift velocity generally increases with $E$. Figure~\ref{fig:vd_Ep} shows the drift velocity as a function of the pressure-scaled $E$ for the alkane concentrations  10\%, 50\%, and 100\% (90\% for \propane).  In particular, at $\Ep<\unit[50]{\uEp}$, $\vd\propto E^a$, where $a\sim1.0\textrm{--}1.2$. In this quasilinear region, the electron mobility~\cite{Rolandi:2008qla},
\begin{align}
    \mob\equiv\frac{\vd}{E},\label{eq:mobility}
\end{align}
is largely field-independent. For pure alkane at \unit[10]{bar} [Figure~\ref{fig:vd_Ep} (c)], the typical mobility is $\sim\unit[0.001]{\umob}$. Furthermore,  the drift velocity variation is $\oom(\%)$ for every \unit[1]{\uEp} change, as can be seen in Figure~\ref{fig:vd_Ep}. Compared to TPCs operating at atmospheric pressure, the pressure variation in a high-pressure TPC is relatively better under control as the pressurized vessel is not connected to the atmosphere.

\subsection{Diffusion} \label{ssec:diffusion}

Once liberated, the primary ionization electrons start to diffuse in all directions through scattering on gas molecules.  The size of the spread grows with time $t$ as $\sim\sqrt{t}$. Under the influence of an electric  field, the diffusing electron clouds drift and the spread in the transverse and longitudinal direction to the field are characterized by $\dlt\sqrt{\ldrift}$, where \dt and \dl are the transverse and longitudinal diffusion coefficients, respectively, and $\ldrift=\vd t$ is the drift length. Diffusion limits the TPC point resolution and track separation threshold. For its momentum reconstruction in a high-multiplicity environment, ALICE chose \mbox{$\dt=\dl=\unit[220]{\uDiff}$}~\cite{Dellacasa:2000bm}. In T2K, the near detector TPC has $\dt=\unit[265]{\uDiff}$~\cite{Abgrall:2010hi, Abe:2019whr}.

\begin{figure}[bth]
    \centering
    \includegraphics[width=\columnwidth]{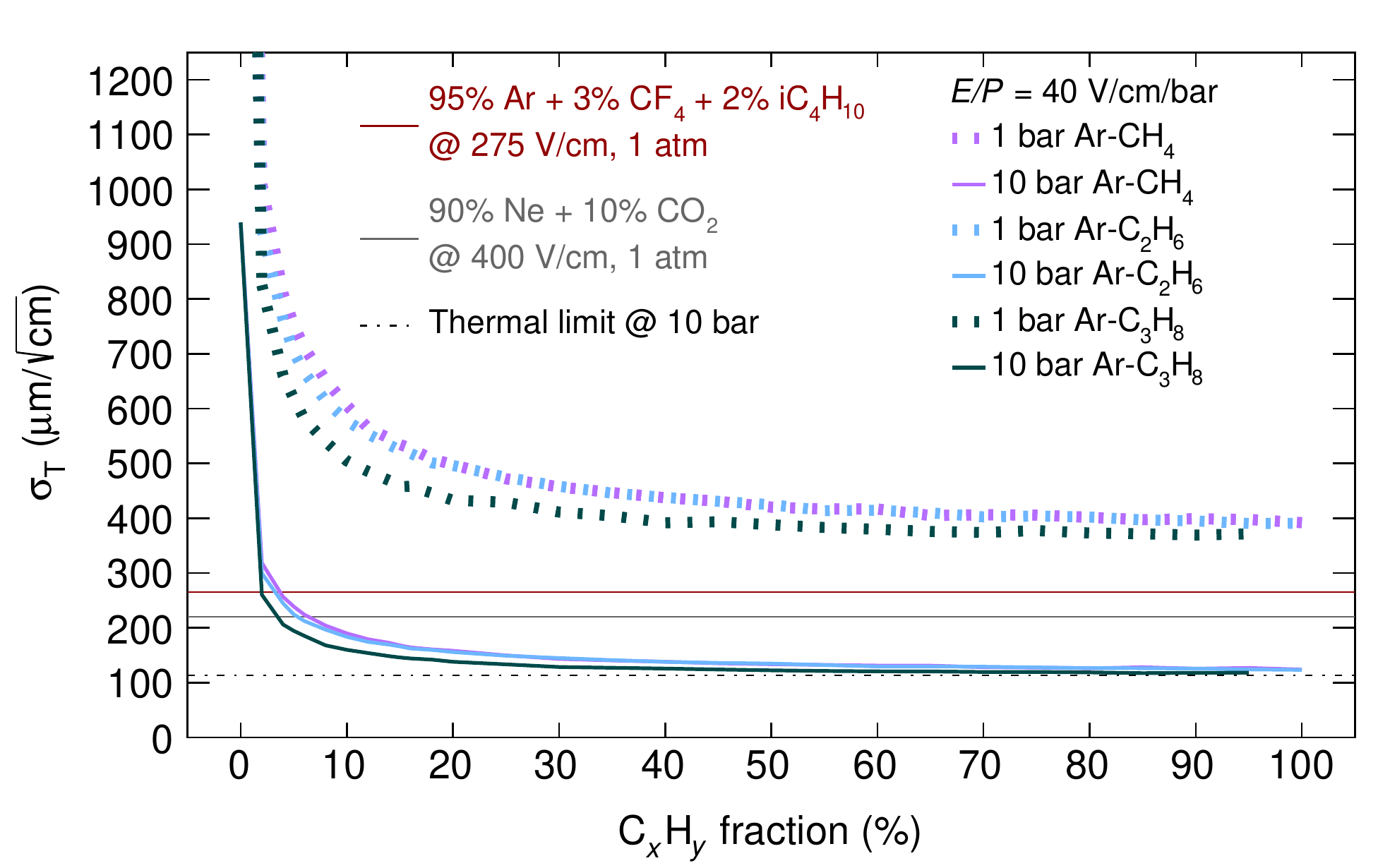} 
    \caption{Transverse diffusion coefficient calculated at \unit[40]{\uEp} as a function of the alkane concentration. The reference values from ALICE and T2K and the thermal limit  for \unit[10]{bar} are shown.}
    \label{fig:pt_fquencher}
\end{figure}

\begin{figure}[!t]
    \centering
    \includegraphics[width=\columnwidth]{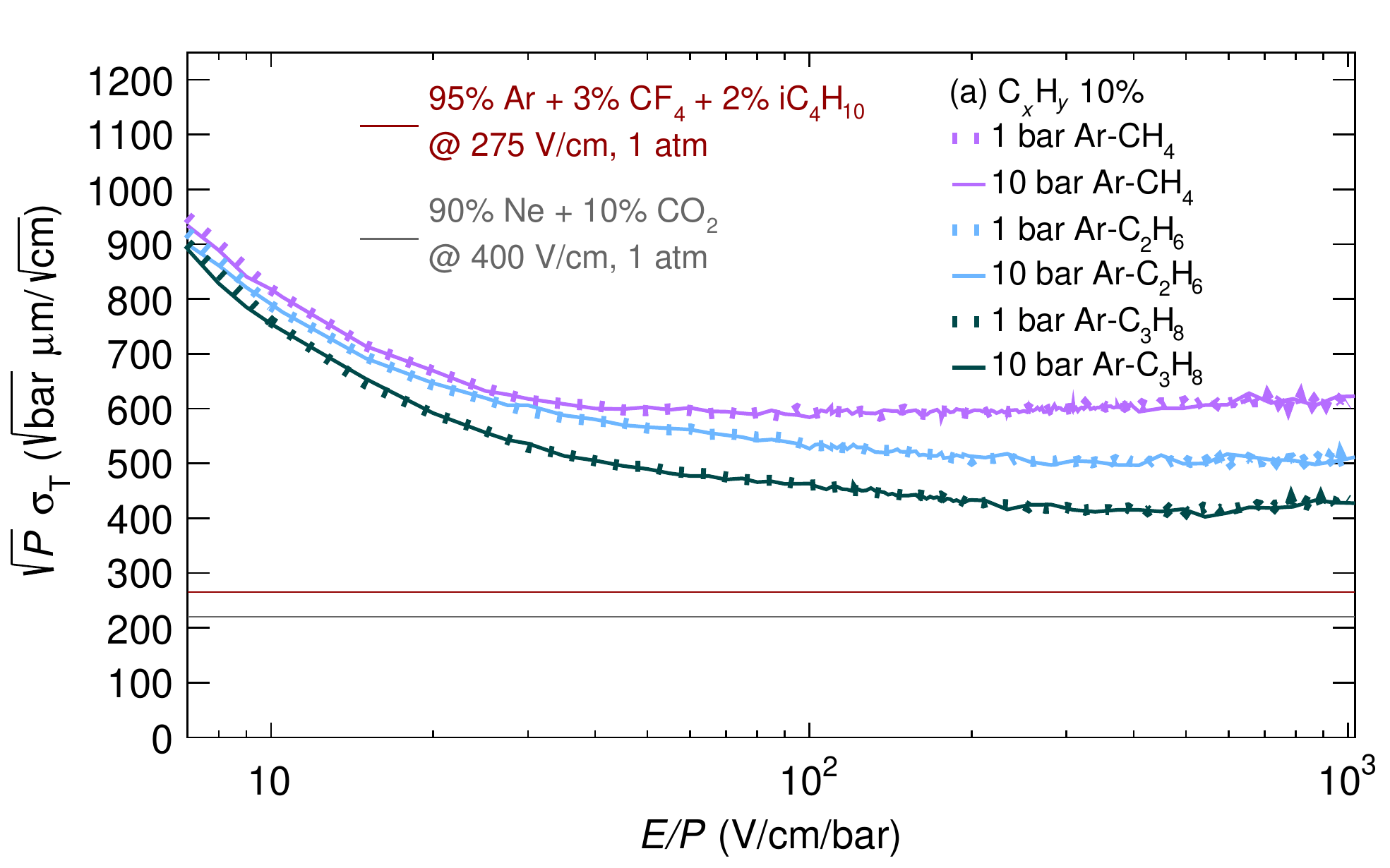} 
    \includegraphics[width=\columnwidth]{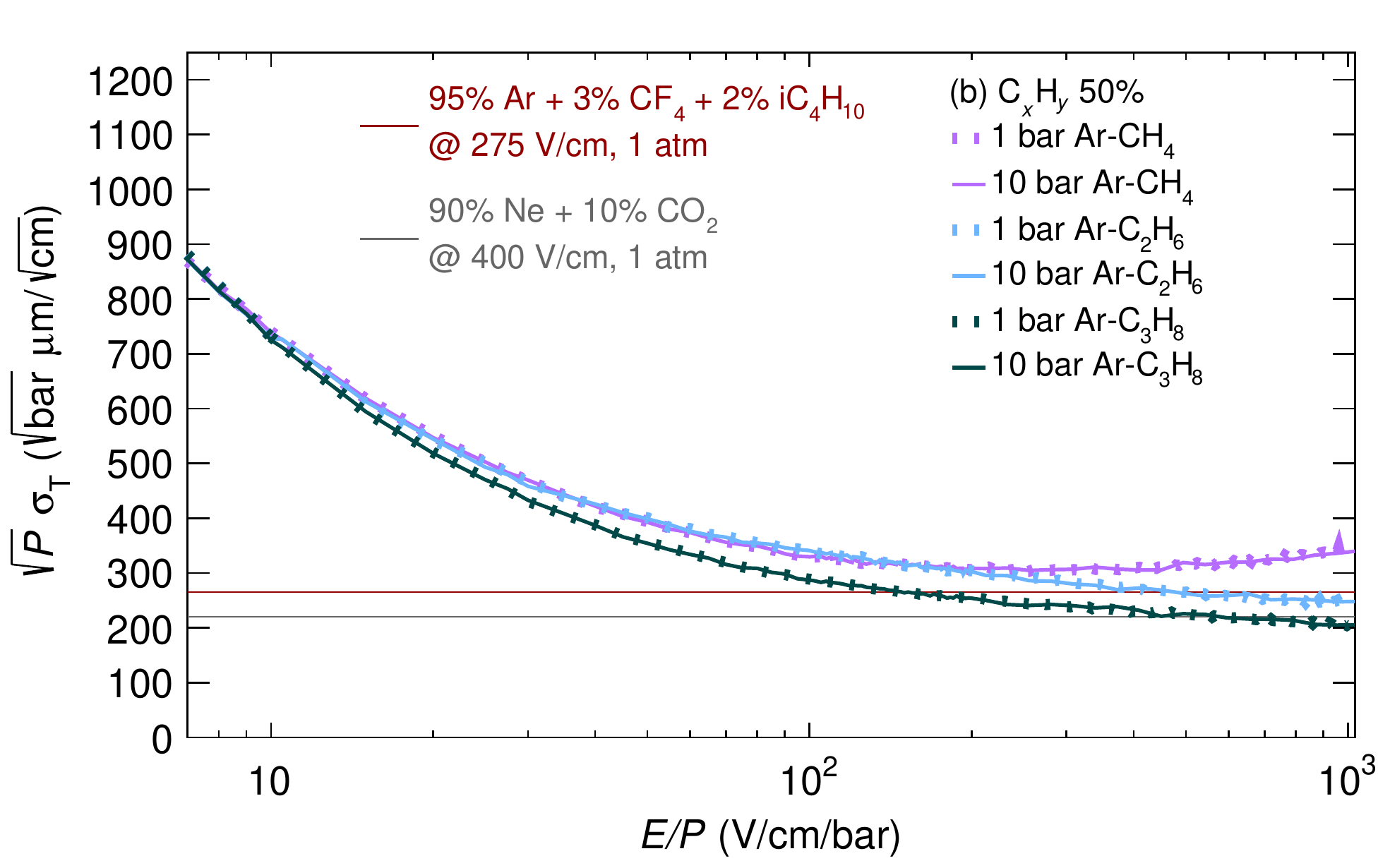} 
    \includegraphics[width=\columnwidth]{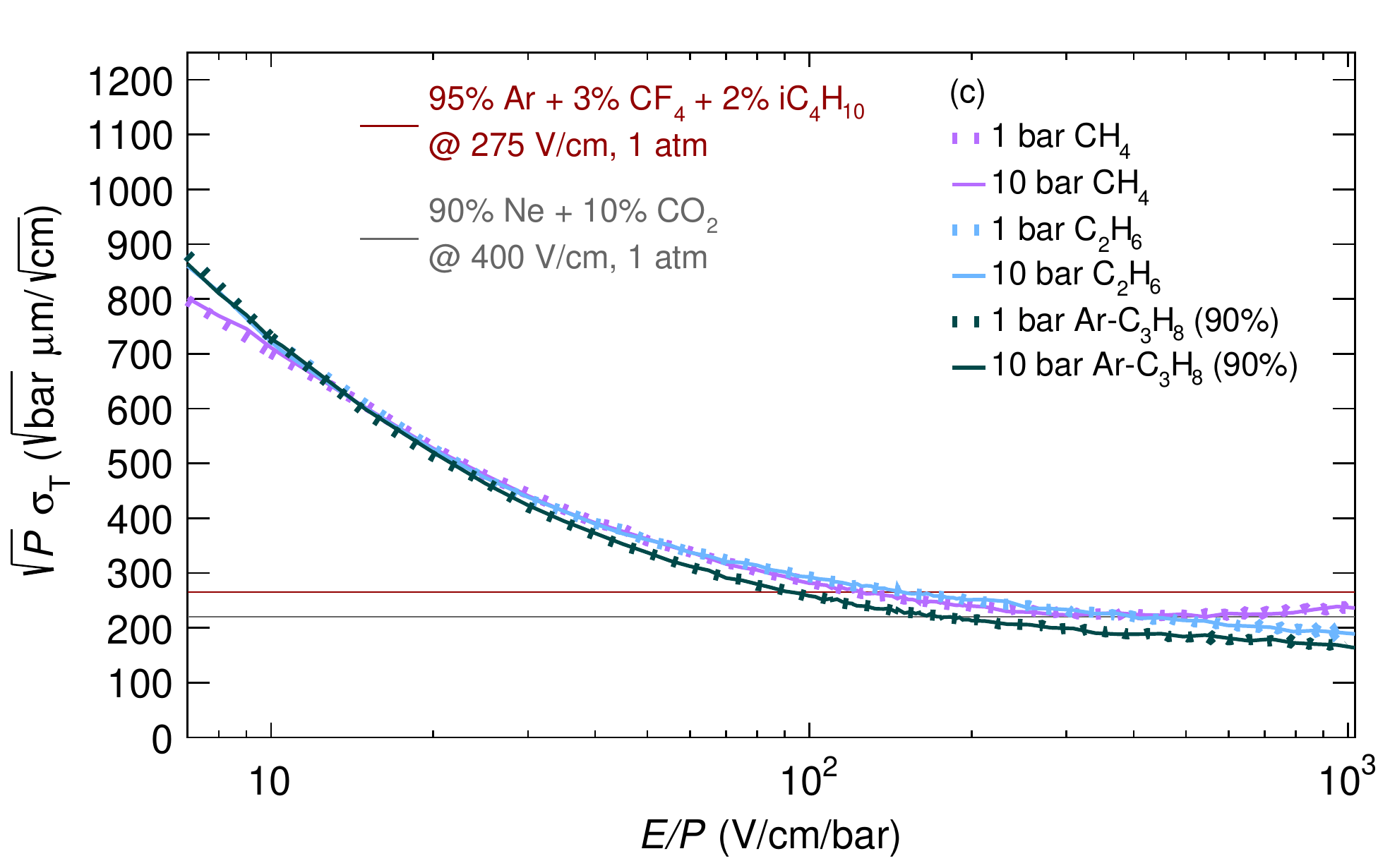} 
    \caption{Scaled transverse diffusion coefficient as a function of the scaled electric field strength. Note the $\sqrt{\pres}\dt(\Ep)$-scaling (see also Table~\ref{tab:density_corrections}). The reference values from ALICE and T2K are shown.}
    \label{fig:pt_Ep}
\end{figure}

The calculated \dt for various Ar-alkane mixtures at \unit[40]{\uEp} is shown in Figure~\ref{fig:pt_fquencher}. Because of the $1/\sqrt{\pres}$-suppression at the same \Ep (Table~\ref{tab:density_corrections}), the diffusion in \unit[10]{bar} for most of the mixtures is smaller than in ALICE---nearly by half for concentrations above \unit[20]{\%}. It   slowly decreases with the  alkane concentration and approaches the thermal limit at \unit[113]{\uDiff} for \Ep=\unit[40]{\uEp}~\cite{Rolandi:2008qla}. In Figure~\ref{fig:pt_Ep}, the transverse diffusion is shown to decrease  with $E$ in the practical region \unit[40--200]{\uEp} (except for \ppten, where it becomes stable). In addition, at  higher concentration, as it approaches the thermal limit, the dependence on $E$ of different alkane becomes similar.

The longitudinal diffusion was also calculated and shows similar size and trends as the transverse diffusion. A comparison between the two at \unit[10]{bar} and  \unit[40]{\uEp} is shown in Table~\ref{tab:diff_10_100}.

\begin{table}[!t]
    \centering
    \begin{ruledtabular}
        \begin{tabular}{lccc}
                & Concentration (\%)         & \dl (\uDiff) & \dt (\uDiff) \\
            \hline
            \multirow{2}{*}{\methane}   & \unit[10]{}     & \unit[224]{}    & \unit[190]{} \\
                                        & \unit[50]{}     & \unit[174]{}    & \unit[133]{} \\
                                        & \unit[100]{}    & \unit[151]{}    & \unit[124]{} \\
            \hline 
            \multirow{2}{*}{\ethane}    & \unit[10]{}     & \unit[189]{}    & \unit[183]{} \\
                                        & \unit[50]{}     & \unit[150]{}    & \unit[134]{} \\
                                        & \unit[100]{}    & \unit[133]{}    & \unit[123]{} \\
            \hline
            \multirow{2}{*}{\propane}   & \unit[10]{}     & \unit[174]{}    & \unit[160]{} \\
                                        & \unit[50]{}     & \unit[130]{}    & \unit[122]{} \\
                                        & \unit[90]{}     & \unit[122]{}    & \unit[118]{} \\
        \end{tabular}
    \end{ruledtabular}
    \caption{        Longitudinal and transverse diffusion coefficients for  Ar-alkane mixtures calculated at \unit[10]{bar} and  \unit[40]{\uEp}.
    }
    \label{tab:diff_10_100}
\end{table}

With an additional  magnetic field parallel to the electric field, $\vec{B}\,||\,\vec{E}$, the transverse diffusion is suppressed~\cite{Rolandi:2008qla}:
\begin{equation}
    \frac{ \dt(B) }{ \dt(0) } = \frac{1}{\sqrt{1+(B\mob)^2}},
\end{equation}
 where $\mob$ is the electron mobility [Eq.~(\ref{eq:mobility})]. 
 For pure alkane at \unit[10]{bar} where $\mob\sim\unit[0.001]{\umob}$ (see Section~\ref{ssec:vd}),  the suppression  by a  \unit[0.5]{T} magnetic field is less than 1\%. Furthermore, with the mobility at \unit[1]{bar} $\sim\unit[0.01]{\umob}$, the magnetic field required to produce the same suppression achieved at \unit[10]{bar} is $B=3/\mob=\unit[3]{T}$. It is interesting to note that, for the longitudinal diffusion,  while it can also be reduced by pressure scaling, it is not affected by the parallel magnetic field.
 
\subsection{Gas Gain} \label{ssec:gain}

After propagation through the drift region, electrons are multiplied in strong electric fields that start ionization avalanches.
Electrodes pick up the amplified signal which can then be more easily digitized by a number of electronics.
In an amplification region with a spatial coordinate $s$ ($s_0<s<s_1$), the gas gain $G$ depends on the path of the electrons~\cite{Rolandi:2008qla}:
\begin{equation}
    G = \exp\left[\int_{s_0}^{s_1}{\left(\alpha-\eta\right)}\mathrm{d}s\right],
    \label{eqn:gain}
\end{equation}
where $\alpha$ is the first Townsend coefficient and $\eta$ the attachment coefficient, both being functions of the electric field strength $E(s)$. 
In our calculation, there are no impurities in the Ar-alkane mixtures, so attachment can not proceed via three-body processes~\cite{Rolandi:2008qla}.
Generally, the attachment is a small correction to the amplification; however, at amplification onset, the attachment cannot be neglected and the effective Townsend coefficient $\alpha-\eta$ is considered.
There is no attachment for the fields relevant in the drift region.

Following the density correction in Table~\ref{tab:density_corrections}, the effective Townsend coefficient is enhanced by the pressure but delayed in onset field due to a shortened electron mean free path---a larger field strength is needed to provide enough energy to initiate the avalanche.

The calculations in this work consider Penning transfer contributions to $\alpha$.
Penning transfers are ionizing energy transfers between gas molecules or atoms and can be summarized by a single transfer coefficient \penningprob that enhances $\alpha$~\cite{Sahin:2010ssz}, in which case $\alpha$ in Eq.~(\ref{eqn:gain}) is replaced by $(1+\penningprob)\alpha$.
Values for \penningprob have been calculated from gas gain measurements for some common argon-based drift gases, but not for quencher fractions above \unit[10]{\%}~\cite{Sahin:2010ssz}.

\begin{figure}[!t]
    \centering
    \includegraphics[width=\columnwidth]{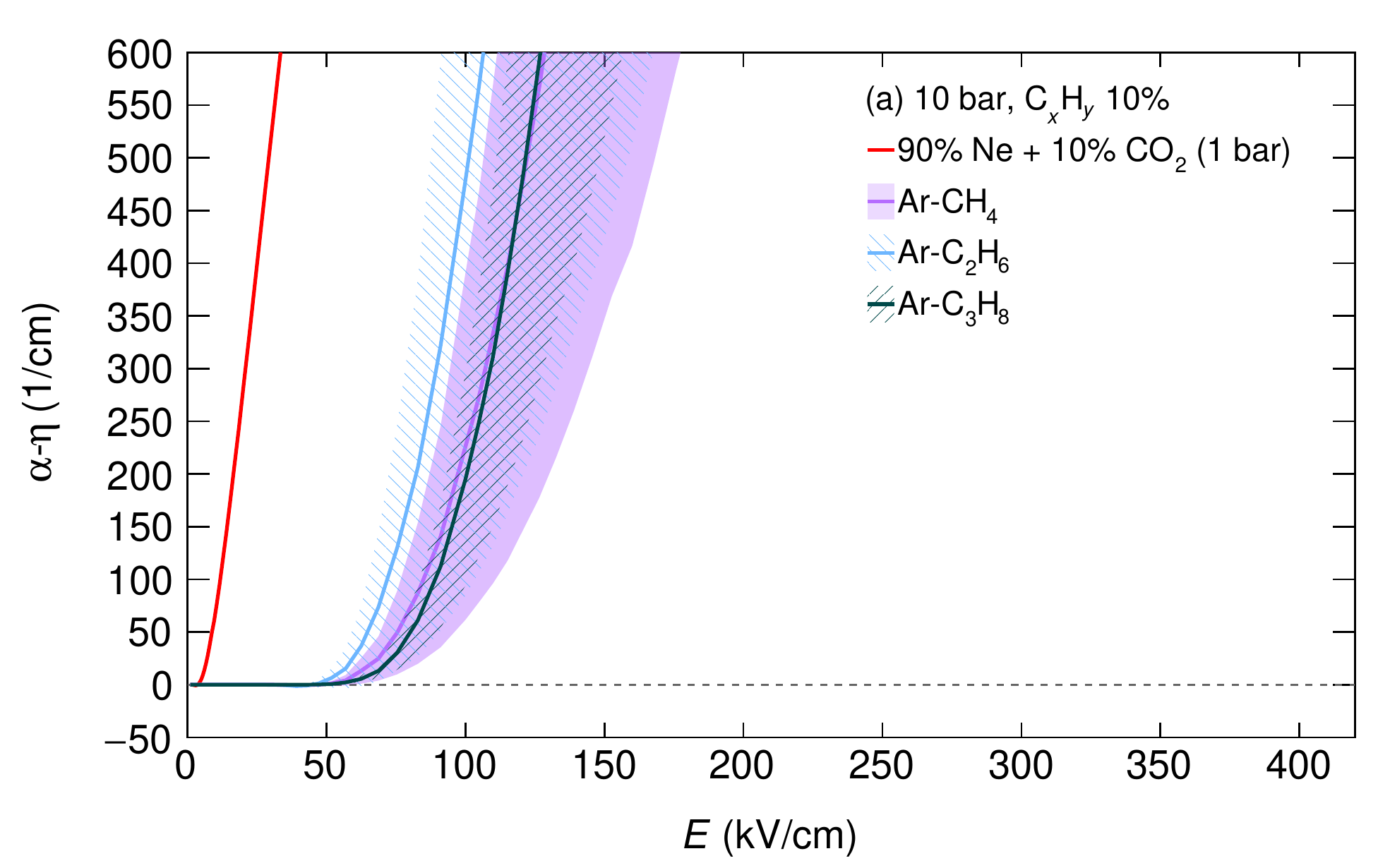}    \includegraphics[width=\columnwidth]{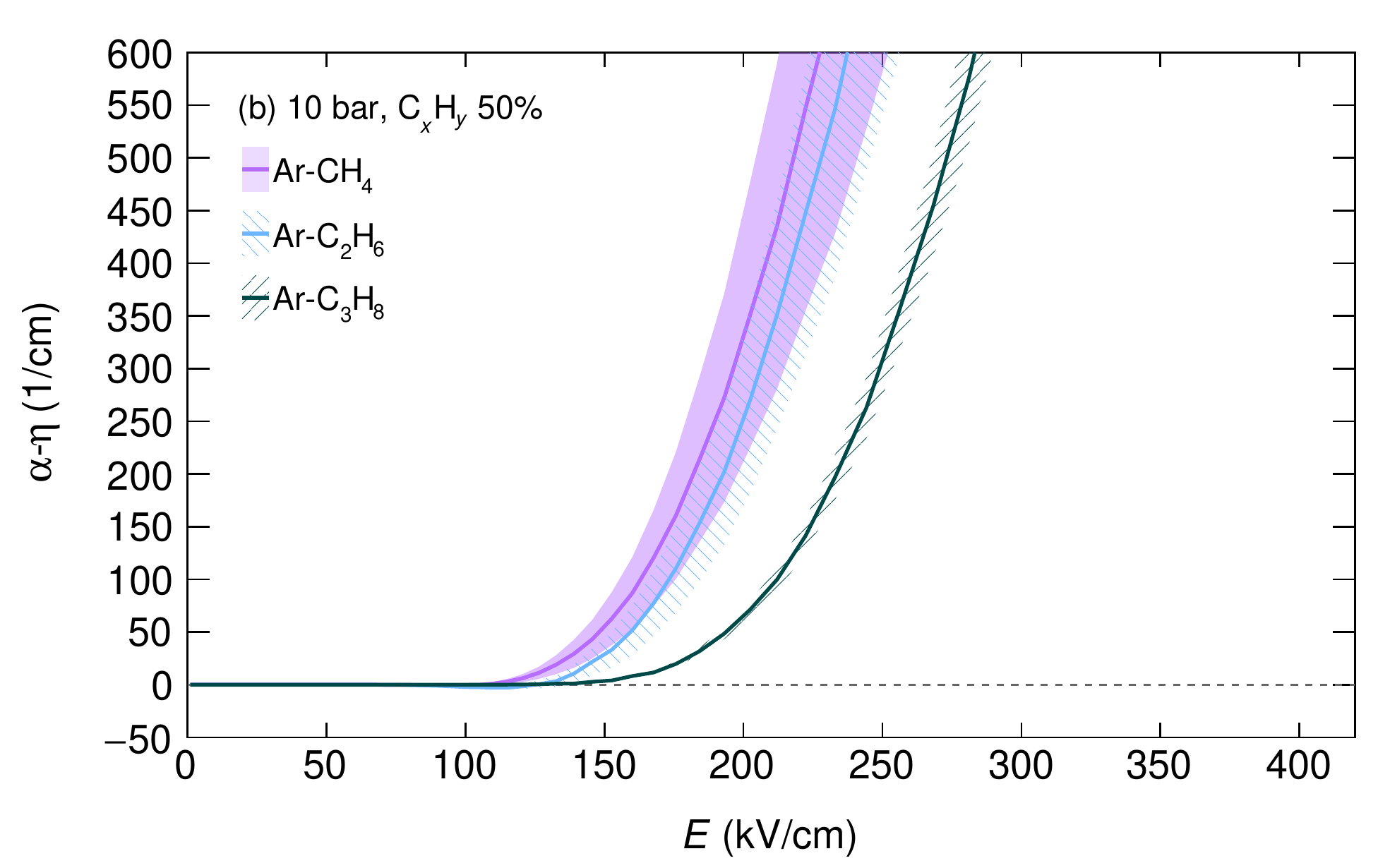}
    \includegraphics[width=\columnwidth]{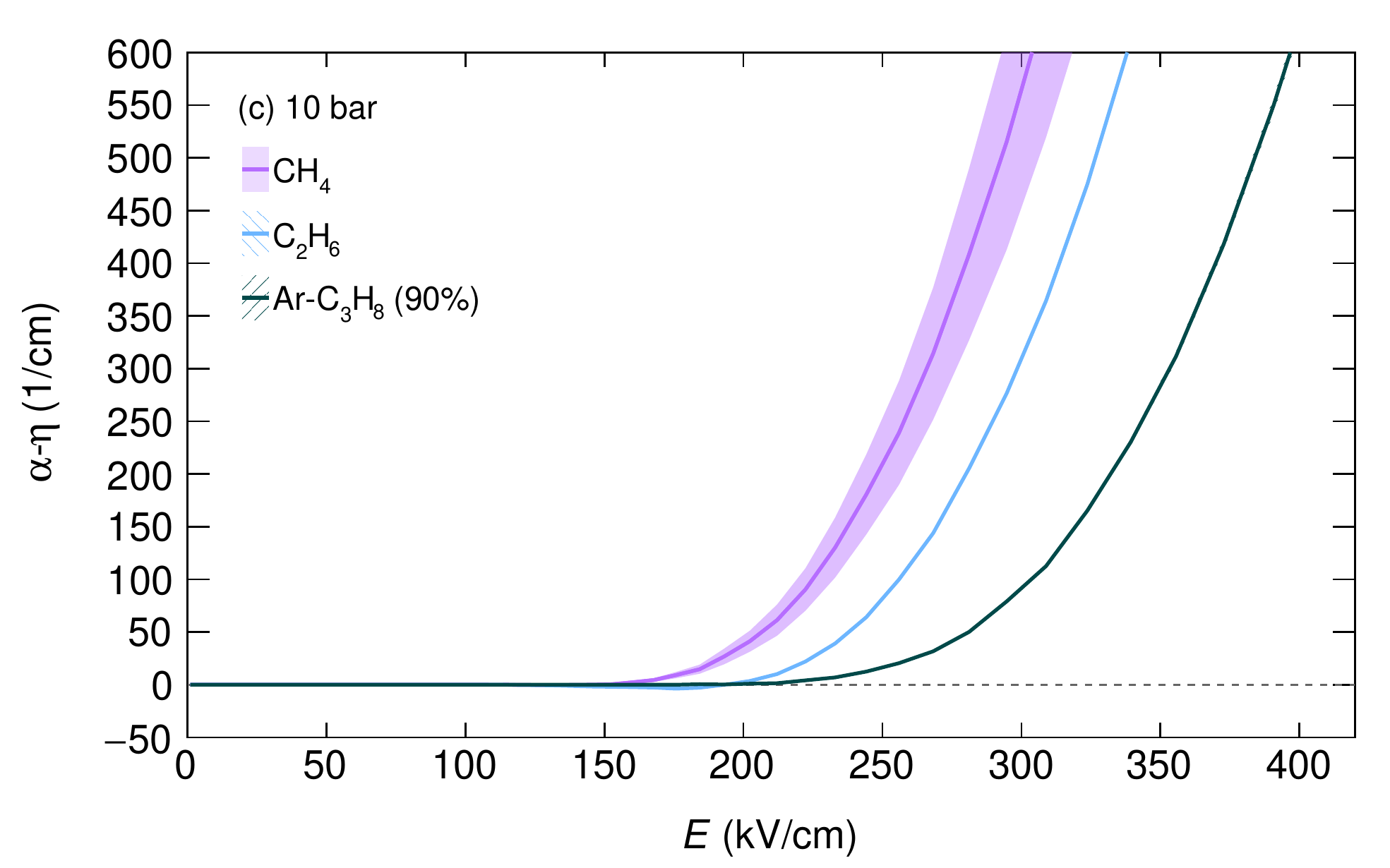}
    \caption{
        Effective Townsend coefficient calculated at \unit[10]{bar} as a function of the electric field strength. The dashed line is to guide the eye for $\alpha-\eta=0$. 
        In panel (a), the simulated values for \alicegas at \unit[1]{bar} (\penningprob determined from~\cite{Sahin:2016bar}) are shown for comparison.  
        The error bands for the Ar-alkane mixtures show the effect of varying \penningprob from \unit[0]{\%} (lower bound) to \unit[100]{\%} (upper bound), while the lines correspond to $\penningprob=\unit[50]{\%}$. 
        In panel (c), the \ethane curve has no error band due to the absence of Penning transfer, whereas the Ar-\propane curve has an error band that is too small to be visible. 
    }
    \label{fig:alpha_eta_fquencher}
\end{figure}

The calculated effective Townsend coefficients for different Ar-alkane mixtures at \unit[10]{bar} are shown in Figure~\ref{fig:alpha_eta_fquencher}.
For Ar-alkane mixtures, the difference between no and maximal \penningprob can be very large, especially for \methane, but is expected to decrease with quencher fraction as transfers from excited \argon to \cphq become less frequent.
It is interesting to note that, pure methane has excited states that exceed the ionization threshold of the molecule, opening the possibility for ionizing energy transfers within the quenching gas itself~\cite{Sahin:2010ssz} [Figure~\ref{fig:alpha_eta_fquencher} (c)]. 

Due to the \Ep-scaling, the onset field strength at \unit[10]{bar} is a factor of 10 larger than at \unit[1]{bar}, as exemplified by the comparison with the ALICE TPC gas (\alicegas, \unit[1]{bar}) in Figure~\ref{fig:alpha_eta_fquencher} (a).
The onset is shown to increase along with the alkane concentration, from \unit[50]{\ukiloE} at \unit[10]{\%} to \unit[150--250]{\ukiloE} for pure alkanes.
High concentrations of propane require significantly larger amplification fields to reach $\alpha-\eta$ values comparable to methane and ethane. In comparison, the gain onset for the ALICE TPC gas is at around \unit[4]{\ukiloE}~\cite{Dellacasa:2000bm, ALICE:2014qrd}.

The need for such high voltages poses a challenge to currently existing gas amplification structures. As examples, wire chambers~\cite{Dellacasa:2000bm} and \micromegas~\cite{Giomataris:1995fq} are typically operated at about $\unit[10-100]{\ukiloEp}$.
The biasing voltage needed to achieve these fields varies between $\oom(\unit[100-1000]{V})$. At \unit[10]{bar}, the bias voltage would be up to $\oom(\unit[10]{kV})$; a challenge for high voltage safety from spark protection to electrostatic distortion of wires.
The significantly higher voltages needed for high fractions of \propane might prove prohibitive in order to reach sufficient gas gain.
A new technology, the resistive \micromegas, has proven to be operational under such high fields close to $\unit[80]{\ukiloE}$ and \unit[1]{bar}~\cite{Procureur:2012ds}; meanwhile, conventional \micromegas have been successfully used in pressurized TPCs (see, for example, Ref.~\cite{Cebrian:2012sp}).


\section{Summary and Discussions}\label{sec:summary}

In this paper, we study the feasibility of measuring neutrino-hydrogen interactions in a \hptpc using argon-alkane gas mixtures. The charged-particle sensitivity of the TPC and its full acceptance and low threshold  make it ideal for a measurement of the neutrino exclusive  $\mu\proton\pi$ production [Eqs.~(\ref{eq:nures}) and~(\ref{eq:nubarres})] that could be used to identify interactions on the hydrogen component out of other nuclear target backgrounds~\cite{Lu:2015hea}. With event-generator calculations, we confirm the efficient phase-space coverage of the detector. By modeling the detector response to the \dptt observable [Eqs.~(\ref{eq:topo}) and~(\ref{eq:sdptt}) ], we demonstrate that the signal-background ratio could be efficiently enhanced by improving the tracking resolution. A hydrogen-enriched TPC gas in an $\oom(\unit[100]{m^3})$ volume at \unit[10]{bar} could not only further increase the signal-background ratio but also deliver a significant event rate. Methane, for example, would provide $\sim10^4$ event per year (at the projected DUNE exposure) with purity above \unit[90]{\%} (assuming a \dptt resolution of $\Gamma=\unit[5]{\mevc}$). Such an event rate is twice what pure hydrogen would yield. The highest hydrogen mass would be provided by $\propnt$, nearly doubling the signal yield of pure \methane and with a signal-background ratio twice as good as by polystyrene.

We also examine the gas-mixture properties related to TPC tracking. At  high  pressure, the effective drift field is reduced. Due to the limitation on megavolt power supplies, the electric field strength across several meters of  drift length  will not be strong enough to saturate the drift velocity to reach the stable maximum. In the practical operational region ($\Ep\sim\unit[40\textrm{--}200]{\uEp}$) we consider,  the drift velocity is (quasi)linear to the  field strength with electron mobility $\sim\unit[0.001]{\umob}$ at \unit[10]{bar}. The resulting drift time could  comfortably  cope with  the highest  event rates foreseen in future accelerator-neutrino experiments. The sensitive drift velocity poses a constraint on the gas system: a per-mil-level stability requires a per-mil-level control on the gas composition, the drift field strength, as well as the temperature and the pressure. 

The high  pressure also  reduces both transverse and longitudinal diffusion to significantly below the ALICE values. At $\Ep=\unit[40]{\uEp}$ and high alkane  concentration, the diffusion coefficients approach the thermal limit $\sim\unit[100]{\uDiff}$ and become almost independent of the alkane type. The impact on the diffusion by a parallel magnetic field is shown to be negligible due to the small electron mobility. 

One further impact on the gas properties by the high pressure is the much stronger amplification field required for gas gain to set in. For pure alkane the onset field strength  is \unit[150-250]{\ukiloE}, about 50 times of the ALICE value. 
And yet, the increased primary ionization density at high pressure (see Appendix~\ref{sec:primaryion}) reduces the overall gain needed to produce usable signals.

In this work, we used the GiBUU neutrino-event generator for the calculation of the signal and background rates.  The underlying nuclear effects  belong to a currently very active research area and the resulting uncertainties on our estimation 
need to be addressed both theoretically and most importantly by dedicated experiments. We modeled the detector response to the TKI observable by a one-parameter smearing function. For the next order accuracy, a  detailed tracking model (and eventually a full detector simulation taking into account the detector geometry) could be applied to the particle-by-particle momentum vectors given by an event generator.  Having these potential future improvements in mind,   we emphasize in this paper the scaling behavior of the signal and background with the tracking resolution and the hydrogen content of the gas. We argue that with the state-of-the-art tracking performance foreseen in a future \hptpc (see Appendix~\ref{sec:multis} for further discussions) and the existing hydrogen-rich gas mixtures,  
neutrino-nucleus interaction background could be reduced to less than \unit[10]{\%}.
It would be crucial at this early stage to estimate a more realistic  $\Gamma$ value that a near-future \hptpc could achieve, 
with the help of detailed detector simulations, 
so that further discussions  could proceed on the physics opportunities provided by  a high-purity neutrino-hydrogen sample. 

In the search of hydrogen-rich gas, we start with argon-hydrocarbon mixtures. The main purpose of the argon component is to provide early-stage synergy with the DUNE argon program. For example, the first-stage hydrogen program could proceed with \ppft to establish the baseline performance while still providing high-statistics  neutrino-argon events (the carbon background might need to be constrained  or statistically subtracted with the help of auxiliary measurements). Except for this practical concern, the argon component could be replaced by helium (see Appendix~\ref{sec:he}), for example, to study neutrino interactions on light nuclei.  The carbon base, on the other hand, is motivated by its small number of (bound) protons. In addition, hydrocarbon, in particular alkane, is a well-studied  TPC gas. As is shown in this work, the drift and gas gain properties with high-concentration alkane do not raise serious concerns in the TPC design. Yet, as it has been mentioned in this paper, the existing calculation of the gas properties could be further improved. In addition to  developing better models for higher order accuracy, dedicated measurements of the gas properties are valuable. It is important to point out that, while we have demonstrated the gas-searching strategy, the gas mixture candidates that we discussed can be further improved, for example, by adding a third component: the ternary mixture Ar-\isobutane-\propane with the isobutane fraction up to \unit[\maxisobutane]{\%} can provide even more hydrogen events than by the Ar-\propane mixture. In fact, if we only consider alkane that is gaseous at \unit[1]{bar}, the theoretical limit for maximal hydrogen mass at \unit[10]{bar} is reached by a mixture that is equivalent to C$_{3.93}$H$_{9.86}$: \unit[17]{\%} \neopentane (neopentane), \unit[35]{\%} \isobutane, \unit[24]{\%} \butane (butane), and \unit[24]{\%} \propane (the first three fractions are determined by the respective vapor pressure at \unit[10]{bar}). Another prospect would be to include gas components that have additional merits like UV transparency~\cite{Azevedo:2017egv}. In addition, the use of flammable gas such as alkane in underground laboratories requires extra precautions; a successful  search for alternative  hydrogen-rich nonflammable gas mixtures would ease this practical concern.  

Finally, we would like to emphasize the impact on neutrino oscillation programs by such a hydrogen-rich \hptpc~\cite{Lu:2015hea}. Around the neutrino energy of the first oscillation maximum at DUNE, pion production is the dominant dynamics. The exclusive processes $\smash{\overset{\scalebox{.3}{(}\raisebox{-1.7pt}{--}\scalebox{.3}{)}}{\nu}}\proton\rightarrow\mu^\mp\proton\pi^\pm$ [Eqs.~(\ref{eq:nures}) and~(\ref{eq:nubarres})], which the hydrogen-extraction technique relies on, could provide constraints to both the neutrino and antineutrino fluxes. The symmetric final states between the neutrino and antineutrino interactions might provide further experimental advantages over the highly asymmetric quasi-elastic dynamics. Measurements of pion production on nucleons also provide critical input for the study of nuclear effects in neutrino-nucleus interactions. 
In addition, there are several possible physics opportunities beyond the oscillation program, as a hydrogen-rich \hptpc revives the possibility of neutrino-hydrogen interaction measurements after 30 years. The processes, Eqs.~(\ref{eq:nures}) and~(\ref{eq:nubarres}), are  the ideal channels to study neutrino deeply virtual meson production ($\nu$DVMP)  where Generalized Parton Distributions (GPDs) could be extracted~\cite{Kopeliovich:2012dr, Siddikov:2019ahb}. Because (anti)neutrinos probe different quark flavors and spins, $\nu$DVMP unfolds the nucleon structure in a complementary way to the GPD program in the proposed Electron-Ion Collider~\cite{Accardi:2012qut}. In addition, because the TKI technique can also be applied to  electron and muon beams---the corresponding leading exclusive channel being $\ell\proton\rightarrow\ell\proton$, where $\ell$ is the electron or muon---electron/muon-hydrogen interactions~\cite{Bernauer:2010wm, Adams:2676885} could be studied by a \hptpc. Furthermore, because of the common detector technology, the extraction technique for $\nu/\ell$-hydrogen interactions could be tested with a small-scale prototype detector at electron/muon beam lines at, for example, Mainz Microtron (MAMI)~\cite{Kaiser:2008zza} or CERN. 

\begin{acknowledgments}
We would like to thank Diego Gonz\'alez-D\'iaz for helpful comments on the manuscript. PH-B would like to thank Stefan Roth for helpful discussions.
PH-B is supported by DFG (Germany) Grant No.\ RO 3625/2-1.
XL is supported by  STFC (United Kingdom) Grant No.\ ST/S003533/1.
JM-A is supported by a fellowship from Fundaci\'on Bancaria ``la Caixa" (ID 100010434), code LCF/BQ/PI19/11690012.
\end{acknowledgments}

\appendix

\section{Multiple Scattering}\label{sec:multis}

The ultimate TPC tracking performance is limited by multiple scattering, diffusion, the geometry of the readout unit, and the field distortion in the drift volume. The first two depend on the gas. As is shown in Section~\ref{ssec:diffusion}, the diffusion is suppressed at high pressure and approaches the thermal limit at high alkane concentration. For completeness, in this section we estimate the size of multiple scattering in the Ar-alkane gas mixtures. 

\begin{figure}[!ht]
    \centering
    \includegraphics[width=\columnwidth]{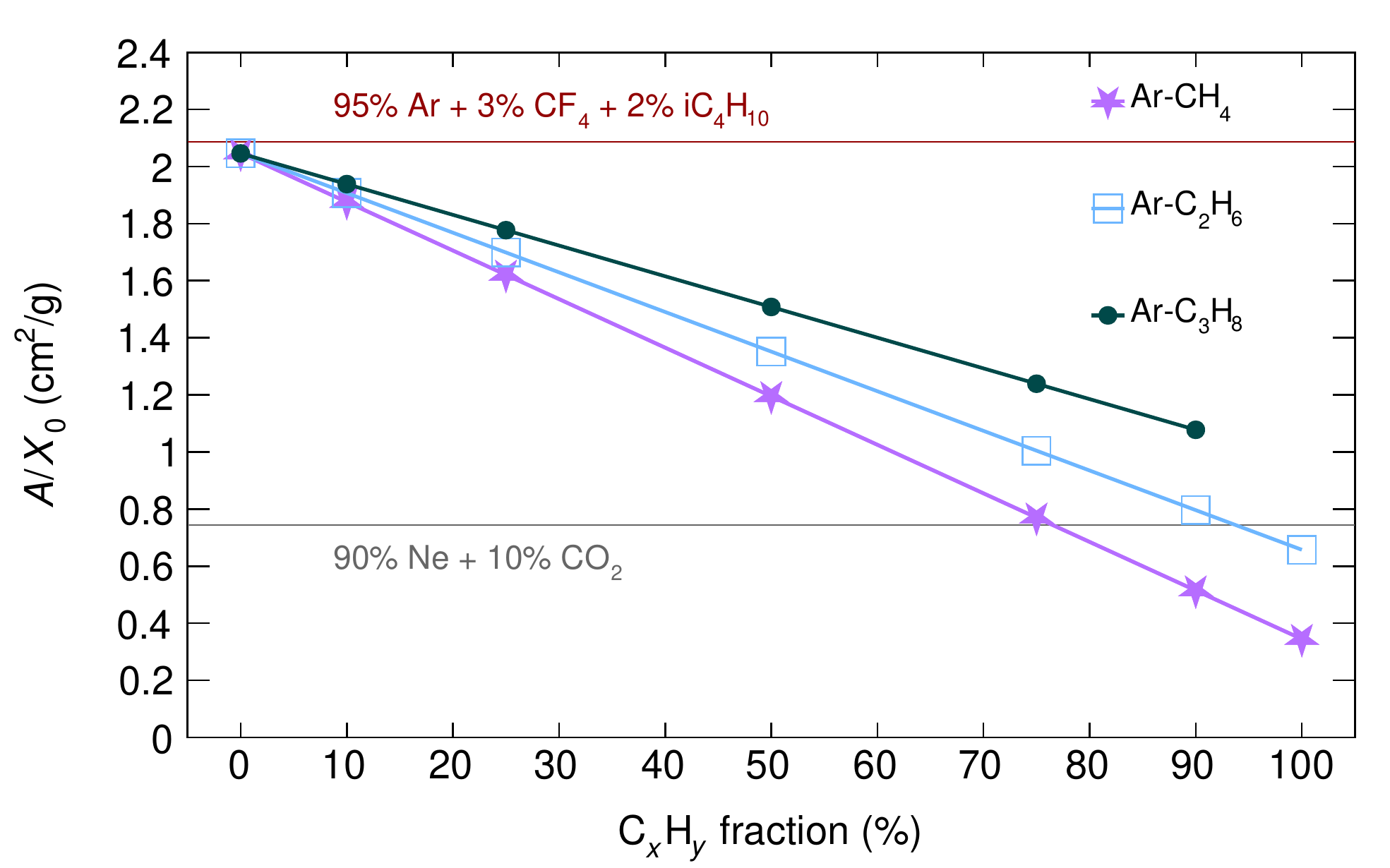}
    \caption{Weighted inverse radiation length, $A/X_0$, as a function of the alkane concentration.  The ALICE (\alicegas) and T2K (\ttokgas) values~\cite{Dellacasa:2000bm} are indicated by the horizontal lines.}
    \label{fig:multis}
\end{figure}

Multiple scattering is commonly quantified by the (r.m.s.) angular deflection, $\theta_\textrm{MS}$. It depends on the radiation length $X_0$ [measured in $\textrm{g}/\textrm{cm}^{2}$ (length$\times$density)]~\cite{Tanabashi:2018oca}:
\begin{align}
\theta_\textrm{MS}=\frac{\unit[13.6]{\mevc}}{p}\sqrt{F}\left(1+0.038\ln F\right),\label{eq:ms}
\end{align}
with the particle momentum $p$ and
\begin{align}
    F=\frac{x}{X_0\beta^2},
\end{align}
where $x/X_0$ is the thickness of material measured in $X_0$ and $\beta$ is the particle velocity in unit of $c$. The gas-dependent part is
\begin{align}
    F\propto\frac{\rho}{X_0}\propto\frac{\pres A}{X_0},
\end{align}
where $\rho$ is the gas density, \pres is the pressure, and $A$ is the atomic mass number. The weighted  $A/X_0$ for different mixtures is shown in Figure~\ref{fig:multis}: it decreases with an increasing alkane concentration. For 10-bar \methane, the $F$ factor is about 4.6 times as large as the one for the 1-bar ALICE gas; the corresponding $\theta_\textrm{MS}$ is therefore about a factor of 2 larger, assuming the log-term in Eq.~(\ref{eq:ms}) is negligible.

\section{Primary Ionization Density}\label{sec:primaryion}

The primary ionization density (\nie) along a track depends on the particle energy lost (\dEdx) and the average energy spent for the creation of an electron-ion pair (\wi).
The value of \wi is a property of the drift gas and independent of the pressure~\cite{Gonzalez-Diaz:2017gxo}, whereas the gas density, and therefore \nie, increases with pressure.
This implies an improvement in the \dEdx measurement with high pressure due to reduced statistical fluctuations.

For the primary ionization density in Ar-alkane mixtures ($\nie^{q}$, $q$ being the quencher fraction), one expects a simple sum rule from the gas components assuming that individual \wi remains unchanged in the mixture:
\begin{align}
    \nie^{q} &= \left[\left(1-q\right)\cdot\nie^{\argon}+ q\cdot\nie^{\cphq}\right]\cdot\pres\left[\mathrm{bar}\right].
    \label{eq:emprical_nei_cphq}
\end{align}
Simulations of minimum-ionizing particles with \heed~\cite{Smirnov:2005yi} are consistent with this expectation, with a deviation smaller than \unit[1]{\%}.
Fitting Eq.~(\ref{eq:emprical_nei_cphq}) to the simulations with various $q$, we obtain 
\begin{align}
\nie^{\argon}&=\unit[26]{\percm},\\
\nie^{\methane}&=\unit[29]{\percm},\\
\nie^{\ethane}&=\unit[48]{\percm},\textrm{~and}\\
\nie^{\propane}&=\unit[59]{\percm}.
\end{align}
However, this simple assumption of unchanged \wi in mixtures is inadequate. Measurements show that \wi in Ar-alkane mixtures is minimal when $q$ lies round \unit[2--3.5]{\%}~\cite{Rolandi:2008qla}.


\section{Helium-Alkane Mixtures}\label{sec:he}

\begin{figure}[!htb]
    \centering
    \includegraphics[width=\columnwidth]{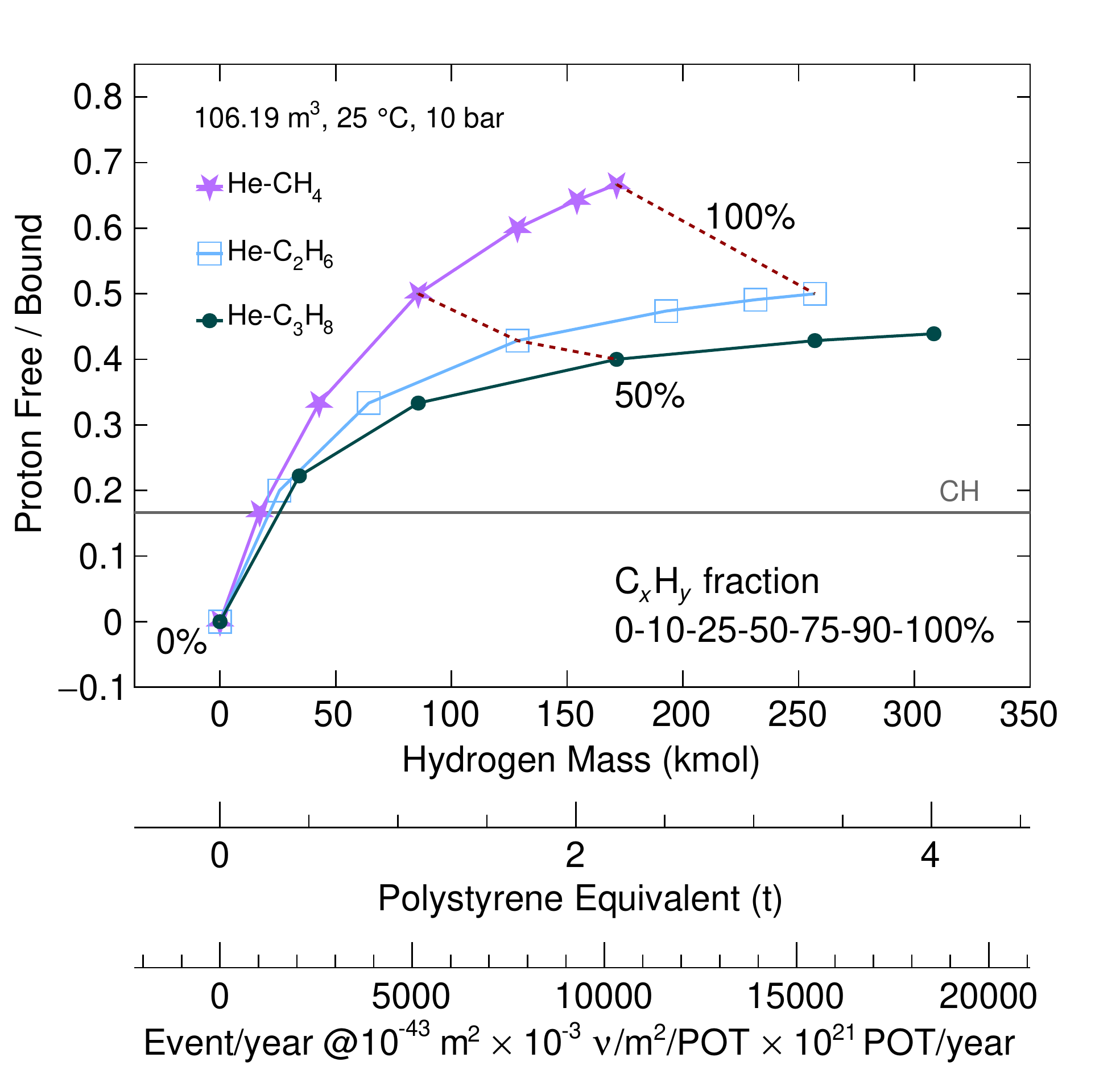} 
    \caption{Proton free-to-bound  ratio vs.\ hydrogen mass for different He-alkane mixtures. See Figure~\ref{fig:hydrogen} for comparison.}
    \label{fig:helium}
\end{figure}

The hydrogen content of  helium-alkane mixtures is summarized in Figure~\ref{fig:helium}. At a given alkane concentration, the hydrogen event rate remains the same regardless of the noble gas (see Figure~\ref{fig:hydrogen} for comparison). On the other hand, at low concentration,  the hydrogen purity is more sensitive to the added alkane because of the small number of bound protons in helium; a high concentration mainly contributes to the increase of the event rate.  

\begin{figure}[!htb]
    \centering
    \includegraphics[width=\columnwidth]{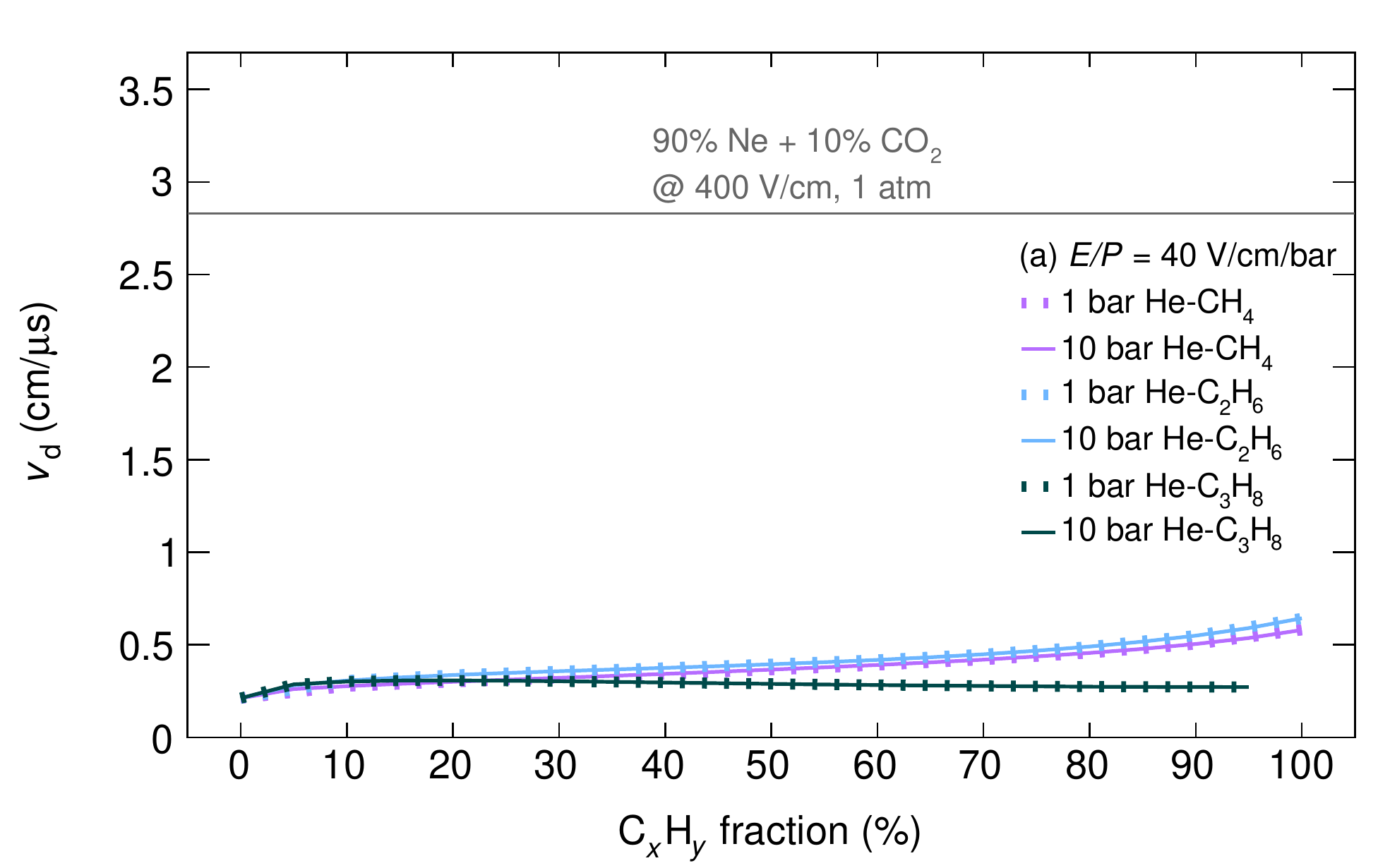}
    \includegraphics[width=\columnwidth]{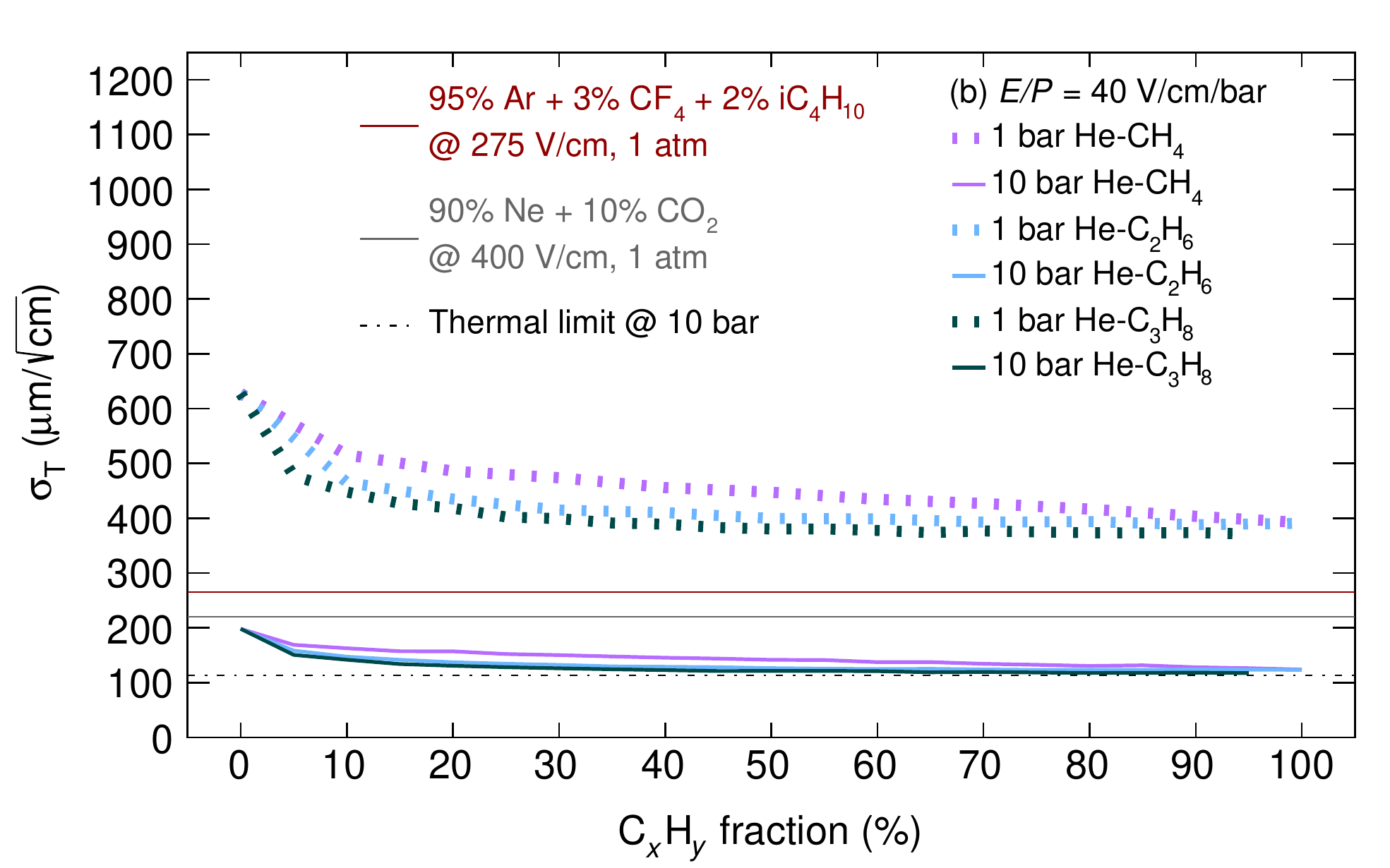}
    \caption{(a) Drift velocity and (b) transverse diffusion coefficient calculated at \unit[40]{\uEp} as a function of the alkane concentration of He-alkane mixtures. See Figures~\ref{fig:vd_cphq_400vcm} and~\ref{fig:pt_fquencher} for comparison.}
    \label{fig:heliumvzxf}
\end{figure}

The calculated drift velocity and transverse diffusion coefficient as a function of the alkane concentration are shown in Figure~\ref{fig:heliumvzxf}. The axis ranges of the figures are chosen to be the same as in Figures~\ref{fig:vd_cphq_400vcm} and~\ref{fig:pt_fquencher} respectively for a direct comparison. With a helium base, the drift velocity increases much more slowly with the added alkane due to the absence of the Ramsauer minimum in helium~\cite{Ramsauer:1921,Rolandi:2008qla}. In pure argon and helium, the drift velocities are similar (both $\sim\unit[0.2]{\uDrift}$), but the transverse diffusion coefficients are very different---for \unit[10]{bar} helium it is about a factor of 5 smaller and already close to the thermal limit.
%

\begin{figure}[!htb]
    \centering
    \includegraphics[width=\columnwidth]{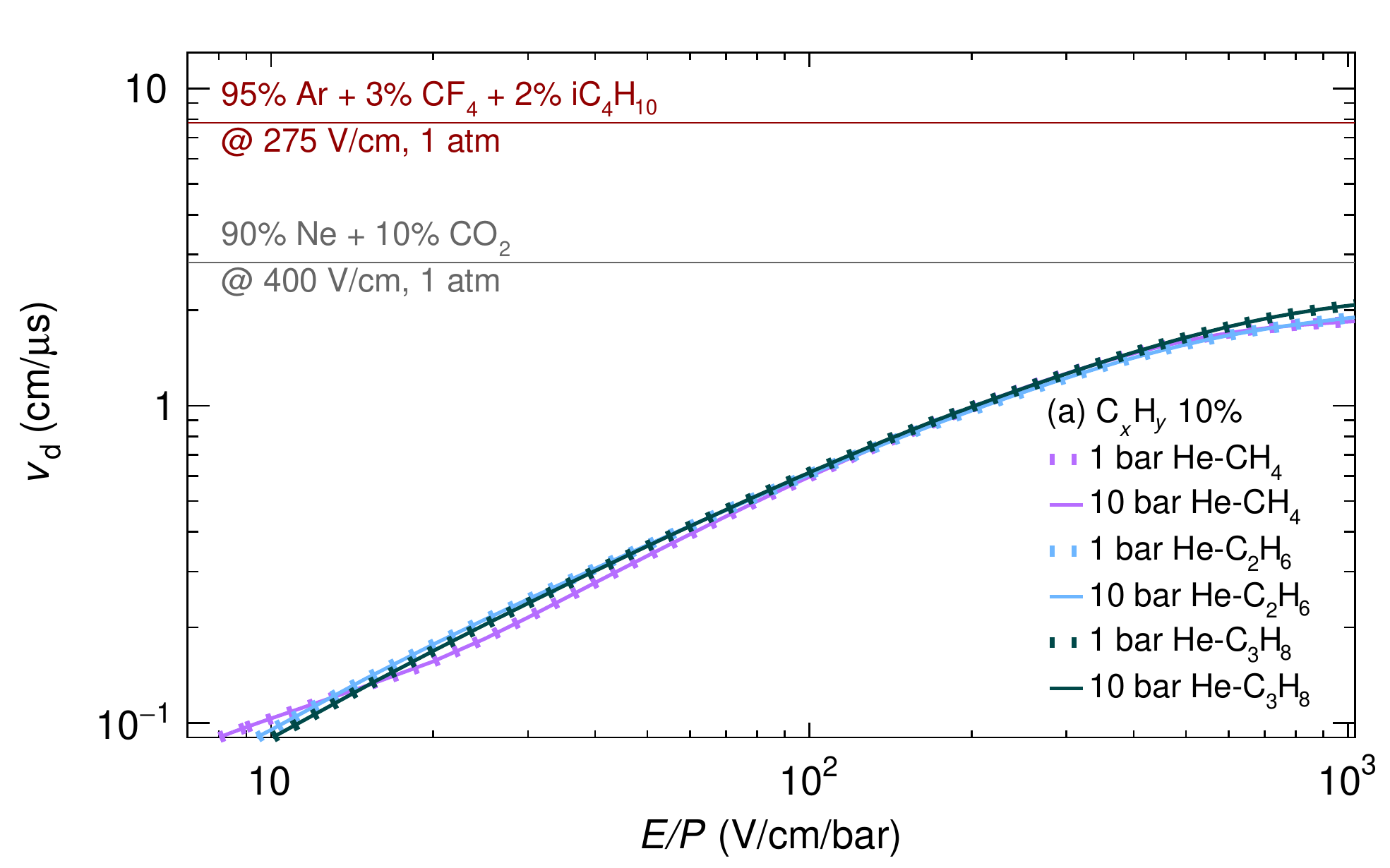} 
    \includegraphics[width=\columnwidth]{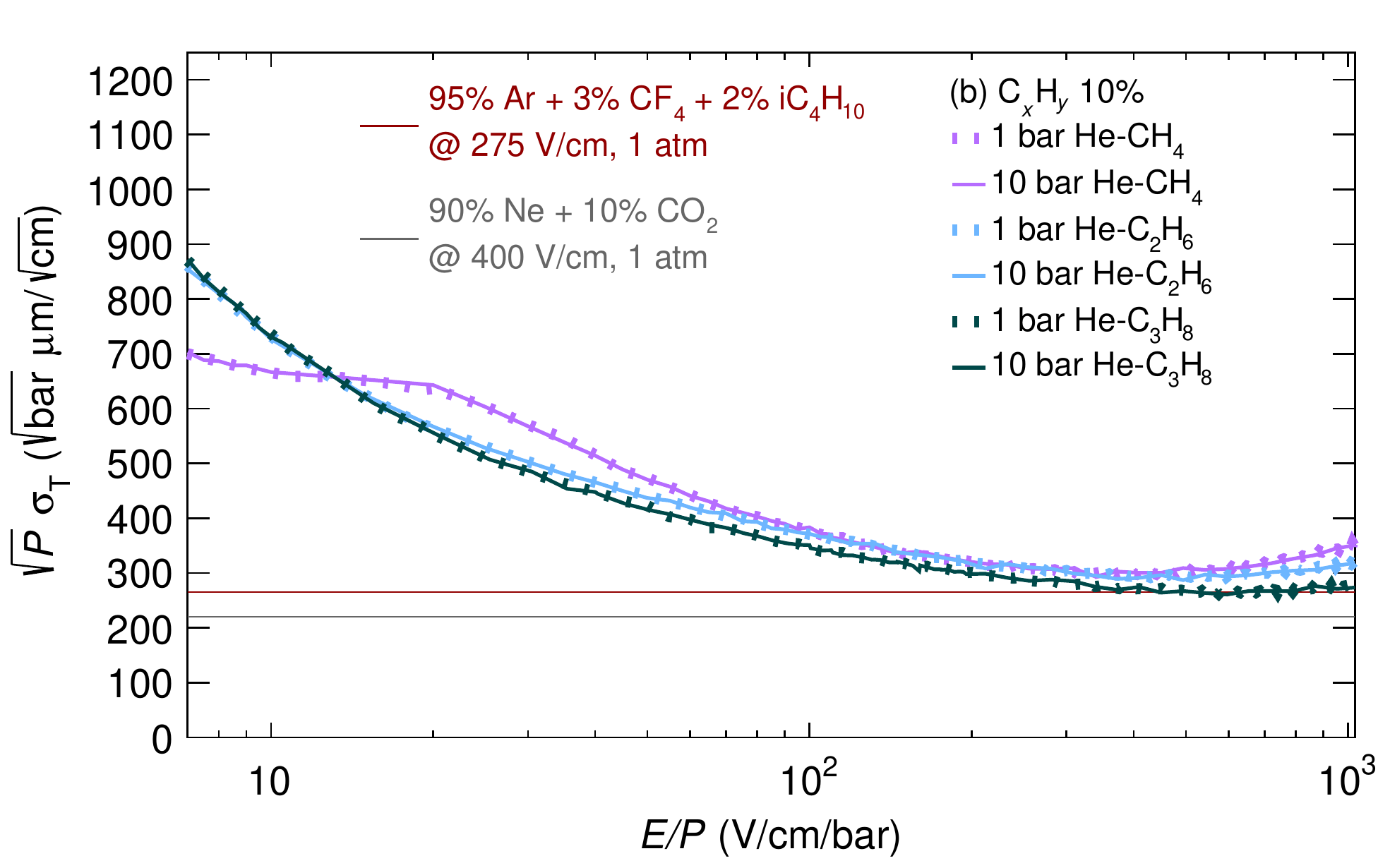} 
    \includegraphics[width=\columnwidth]{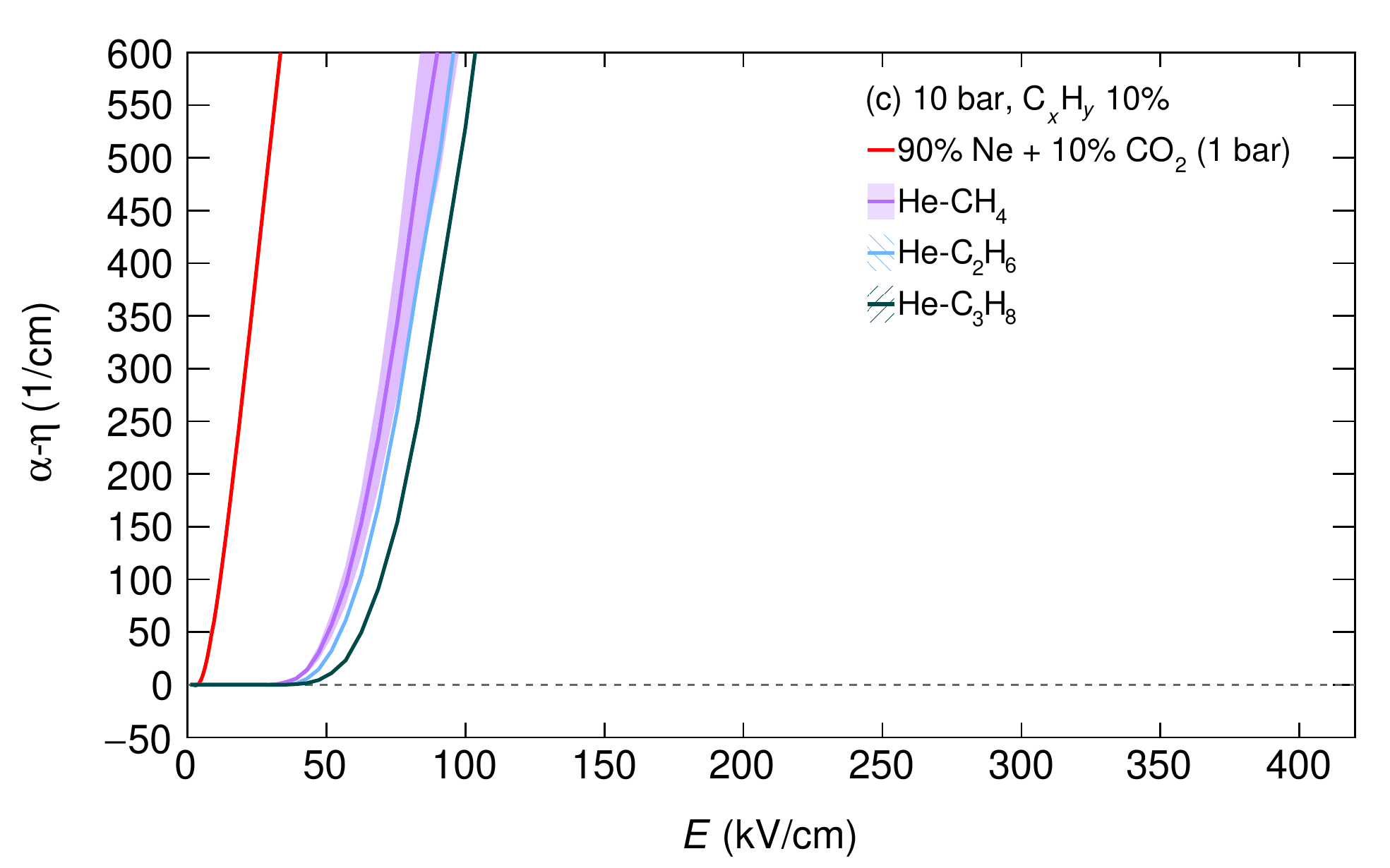} 
    \caption{(a) Drift velocity, (b) scaled transverse diffusion coefficient, and (c) effective Townsend coefficient as a function of the (scaled) electric field strength for He-alkane mixtures at 10\% alkane concentration. See Panels (a) of  Figures~\ref{fig:vd_Ep},~\ref{fig:pt_Ep}, and~\ref{fig:alpha_eta_fquencher} for comparison.}
    \label{fig:heliumall}
\end{figure}

As the difference between helium and argon is more relevant at low alkane concentration, the gas parameters (\vd, \dt, and $\alpha-\eta$) in He-alkane mixtures are shown for the \unit[10]{\%} concentration in Figure~\ref{fig:heliumall}. As discussed previously, because of the absence of the Ramsauer minimum, the drift velocity is significantly smaller across the whole range of $\Ep$. For the transverse diffusion coefficient, the difference between helium and argon is most visible at high $\Ep$. This strong \Ep-dependence is also seen in the gas gain. The gain onset is earlier and much steeper in helium-alkane mixtures. In Figure~\ref{fig:heliumall} (c), Penning transfers always happen for helium mixtures in \magboltz; in the case of \methane, auto-transfers cause the broadening of the curve [see  Figure~\ref{fig:alpha_eta_fquencher} (c) and related discussions].

\bibliography{bibliography}

\end{document}